\documentclass[12pt]{article}
\usepackage{jcappub}
\usepackage{graphicx}
\usepackage{dcolumn}
\usepackage{bm}
\usepackage{amssymb}
\usepackage{amsmath}
\usepackage{color}
\usepackage[latin1]{inputenc}
\usepackage{tikz}
\usepackage{graphics}
\usepackage{upgreek}
\usepackage{enumerate}
\usepackage{eufrak}

\def\be{\begin{equation}}
\def\ee{\end{equation}}
\def\ba{\begin{eqnarray}}
\def\ea{\end{eqnarray}}
\def\beq{\begin{eqnarray}}
\def\eeq{\end{eqnarray}}
\def\ep{\varepsilon}
\def\mpl{M_{\rm Pl}}

\def\d{\mathrm{d}}
\def\L{\mathcal{L}}
\def\E{\mathcal{E}}
\def\({\left(}
\def\){\right)}
\def\ie{{\it i.e. }}

\def\nn{\nonumber}
\def\mn{_{\mu \nu}}
\def\stu{St\"uckelberg }
\def\p{\partial}
\def\mupn{^\mu_{\ \nu}}
\def\<{\langle}
\def\>{\rangle}

\def\D{\mathcal{D}}
\def\K{\mathcal{K}}

\title{Generalized Galileon Duality}
\author{Claudia de Rham$^{a,b}$,}
\author{Luke Keltner$^{a}$,}
\author{Andrew J. Tolley$^{a}$}
\affiliation{$^a$CERCA/Department of Physics, Case Western Reserve University, 10900 Euclid Ave, Cleveland, OH 44106, USA}
\affiliation{$^b$Perimeter Institute for Theoretical Physics, 31 Caroline St N, Waterloo, Ontario, N2L 6B9, Canada\footnote{Emmy Noether Visiting Fellowship}}
\emailAdd{Claudia.deRham@case.edu}
\emailAdd{Lucas.Keltner@case.edu}
\emailAdd{Andrew.J.Tolley@case.edu}

\abstract{We generalize the Galileon duality to any single scalar field Lagrangian coupled locally to any matter field.
Under the duality, a generalized Galileon maps into another generalized Galileon via a one parameter group of transformations, with only a simple modification of the Lagrangian functions. We find a special class of generalized Galileons for which the duality is a symmetry of the action. We further extend the duality to the case of vector fields and give the dual formulations of the Maxwell and Proca theories.
We include arbitrary local couplings to matter fields and show that the duality always maps a local interacting theory into a local interacting theory.
We also discuss the coupling to gravity and uncover a new class of Lorentz invariant massive theories which map into themselves under the duality.
Finally, we show that the duality can be used to map solutions of a theory with superluminal (luminal) group velocity into one with luminal
(subluminal) group velocity.
We find that the duality nevertheless preserves the classical causal structure and emphasize the need to include the quantum corrections to ascertain relativistic causality.
}
\begin{document}
\maketitle
\flushbottom

\section{Introduction}

Dualities are equivalences between naively distinct theories. A duality may be classical - it relates
two distinct classical theories at the level of the classical action usually by means
of a field redefinition, or it may be quantum - the equivalence is only at the level
of the entire path integral or as a map between correlation functions. In either
case it is expected that the (classical) S-matrix remains invariant under the transformation.
In the special case that the duality leaves the action invariant, then it is a symmetry. In the following we shall consider a novel classical duality which we argue is extendible to the quantum level.\\

The `Generalized Galileons' describe the most general class of single scalar field theories in Minkowski spacetime  \cite{Deffayet:2011gz}. They are generalizations of the `Galileon' Lagrangian \cite{Nicolis:2008in} which is the most general Poincar\'e invariant single scalar field theory Lagrangian with second order equations of motion for a field which admits the nonlinearly realized `Galilean' symmetry $\pi \rightarrow \pi + v_{\mu}x^{\mu}$. It was recently observed in \cite{Curtright:2012gx,deRham:2013hsa} that the Galileon Lagrangians admit a nontrivial classical duality which maps one particular class of Galileon models into a second distinct class of Galileon models. We refer to this as a `Galileon duality'.
This duality is closely related to similar results \cite{Creminelli:2013fxa} connecting conformal Galileons and DBI-Galileons \cite{deRham:2010eu} as has been emphasized in \cite{Creminelli:2014zxa}. The duality arises as a twist in the choice of representations of the coset for the Galileon algebra $Gal(3+1,1)/ISO(3,1)$ or conformal algebra $SO(4,2)/ISO(3,1)$ \cite{Creminelli:2014zxa}. In the particular case of the Galileon duality of \cite{Curtright:2012gx,deRham:2013hsa} there is a natural realization of this duality in the context of bi--gravity models \cite{Hassan:2011zd} for which the Galileons arise in a specific decoupling limit \cite{Fasiello:2013woa}. In the bi--gravity picture, the duality is linearly realized, and corresponds to a simple diffeomorphism. Nevertheless in the decoupling limit description, the Galileon appears as an infinite order in derivatives field redefinition which disguises its locality. We shall demonstrate in what follows that in the simpler case of Massive Gravity \cite{deRham:2010kj} the duality is also equivalent to an invertible diffeomorphism, \ie  change of gauge.\\

In this article we will extend the results of \cite{Curtright:2012gx,deRham:2013hsa} to the entire class of Generalized Galileons, \ie all scalar field theories, coupled to matter in a completely general and {\it local} way. We find that the Galileon duality is far more extensive than previously recognized. The duality presented in \cite{deRham:2013hsa}  is equivalent to a Legendre transform of the field variables \cite{Curtright:2012gx}. Here we find it more useful to view the duality transformation as a (field-dependent) diffeomorphism as it arose in \cite{Fasiello:2013woa}. With this in mind, any standard matter field (be it a scalar, vector, tensor, spinor or higher--spin) should transform in the normal way under this diffeomorphism. In particular it maps a Generalized Galileon arbitrarily coupled to matter in a local way, into a distinct dual theory in which matter remains locally coupled to the dual Galileon field. As an extreme example, the entire standard model of particle physics gets mapped to a classically equivalent dual local field theory when the duality transformation is performed on one of the components of the Higgs field. \\

Although the duality we describe works for any scalar field theory, we are particularly interested in the case of models which exhibit the Vainshtein mechanism \cite{Vainshtein:1972sx}. These models are known to have technical issues such as a low strong coupling scale, and classically superluminal group and phase velocities. In particular the Galileon models themselves are examples of theories exhibiting the Vainshtein mechanism. The duality acts nontrivially on these properties, in particular it can be used to map the strong coupling Vainshtein region into a weakly coupled one, or it can be used to map a solution with classical superluminal group velocities into one with subluminal group velocities \cite{Curtright:2012gx,deRham:2013hsa}. This means the duality can be  used as a useful tool to probe the non-perturbative aspects of these theories. It also confirms that classical group velocities are not the real measure of the causality of a theory.  \\

While the Galileons can be defined as scalar field theories with nonlinearly realized symmetries, they are also fields inherited from infrared theories of modified gravity (DGP \cite{Dvali:2000rv}, cascading gravity \cite{deRham:2008zz}, massive gravity \cite{deRham:2010ik}, bi--gravity \cite{Hassan:2011zd}, New Massive Gravity \cite{Bergshoeff:2009hq,deRham:2011ca}, etc.) where the Galileon usually represents the helicity-zero mode of the graviton and is therefore not a scalar under diffeomorphisms\footnote{Only in the decoupling limit when identifying the global Lorentz symmetry with the global subgroup of the local diffeomorphism symmetry does that field behave as a scalar under the combined transformation.}. Instead we shall see in this manuscript that it is the derivative of that field which transforms as a scalar under the duality map. This particular transformation law is at the core of the duality and has many powerful consequences. \\

In what follows we consider a Generalized Galileon coupled to any matter field \cite{Deffayet:2011gz}. Under the duality a local coupling to matter maps into a local coupling to matter. Furthermore since the duality map is associated with a diffeomorphism, it provides a strong indication that the map is also valid at the quantum level (unless diffeomorphism anomalies appear which are known only arise in $4k+2$ dimensions \cite{AlvarezGaume:1983ig}).
We shall find special cases where the duality is itself a continuous global symmetry of the action, which strengthens the case for the validity of the duality at the quantum level for these examples. In particular the pure quintic Galileon in four dimensions is duality invariant. \\

Not only can the Generalized Galileon  couple to any matter field but we also show how to couple the theory to gravity. Massive gravity \cite{deRham:2010kj} naturally appears as an example of a duality-invariant gravitational theory where the duality is simply associated with a new choice of gauge and therefore does not modify the theory.
In addition to massive gravity as presented in \cite{deRham:2010kj} we find a new extension of massive gravity which is also manifestly invariant under the duality map. This new extension of massive gravity is distinct from the master theory\footnote{Although the master theory presented in \cite{Gabadadze:2012tr} is not explored in this manuscript it is extremely likely that this gravitational theory is also invariant under the duality transformation provided that the additional field maps in the appropriate way.} presented in \cite{Gabadadze:2012tr} which involves an additional scalar degree of freedom.  In the extension presented in this manuscript the mass parameters and the Planck scale may depend on the \stu fields themselves. We show that this new theory enjoys the same primary constraint which removes half the Boulware--Deser ghost \cite{Boulware:1973my} as derived in \cite{Hassan:2011hr}. We also show that the secondary constraint is present as in \cite{Hassan:2011ea} which then implies that this extended theory of massive gravity only propagates five degrees of freedom. This theory remains globally Lorentz invariant but breaks translation invariance in the vacuum, and is the reason why it does not take the standard Fierz--Pauli form. \\

We also propose a generalization of the duality involving a global Lorentz vector $A_\mu$. Under this duality map, the Lorentz vector $A_\mu$ transforms as a diffeomorphism scalar. This allows us to find a dual formulation of Maxwell's theory which does not uniquely involve the Maxwell field strength but yet enjoys a non-linearly realized $U(1)$ gauge symmetry and thus only propagates two degrees of freedom in four dimensions. This could open the door for finding new interactions in gauge theories. \\

Finally, as mentioned previously, one of the particularities of the Galileon duality is that it can map a superluminal group velocity into a (sub)luminal one. % In this manuscript we show how the coupling to matter transforms under this duality and
 Following local couplings to matter we
 show that the classical causal structure remains the same in both representations. For instance in one representation the Galileon (or Generalized Galileon) may propagate superluminally and the matter field luminally while in the dual version, the Galileon propagates at the speed of light and the matter field subluminally. While these classical group velocities are not the same, the classical causal structure is the same in both representations, and there is no paradox. We emphasize the fact that the classical velocity does not need to remain invariant under the duality map. Rather it is the front velocity which determines the causal structure of the theory which should remain invariant. Since the front velocity should be computed in the quantum regime, one cannot rely on a purely classical calculation to determine the causal structure. Furthermore the classical superluminal group velocities are not indicative of acausality. A more detailed discussion on how quantum effects could render the front velocity luminal, thus ensuring causality, will be given in \cite{Luke}. \\

The rest of this manuscript is organized as follows. We start by reviewing the duality map in section~\ref{sec:DualityMap}, emphasizing its role as a diffeomorphism and promoting it to a vector duality.  We then apply this duality map to a Generalized Galileon theory in section~\ref{sec:GeneralizedGalileon} and uncover the existence of a class of theories which remain invariant under this transformation. We then discuss how the duality acts on a general vector theory in section~\ref{sec:VectorDuality} and introduce the simple examples of the dual to Maxwell and to a Proca theory.  The general coupling to any matter field is then presented in section~\ref{mattercoupling}. We explain how a local coupling maps into a local one in the dual representation in a way which preserves the information needed on the initial Cauchy surface. We also present an example of a renormalizable theory which admits a local, second representation with irrelevant operators. The coupling to gravity is investigated in section~\ref{sec:Gravity}. We argue that covariant Galileons are not invariant under this map but Massive Gravity is. We also introduce a new extension of Massive Gravity which is also duality--invariant. Finally we address the crucial issue of (super)luminality in this class of theories in section~\ref{sec:SL}. We show how the causal structure is maintained at the classical level despite mapping superluminal group velocities into luminal ones. More importantly however we emphasize the role of the front velocity when determining the causal structure which ought to be computed at the quantum level.  Finally we summarize our results in section~\ref{sec:discussion}.
In Appendix~\ref{sec:appendix} we also show how to promote the duality to a larger class of non-Lorentz invariant ones.

\section{The Duality Map}

\label{sec:DualityMap}

\subsection{The Duality as a coordinate transformation}

The Galileon duality transformations are a one parameter family of invertible field redefinitions. Given a field $\pi(x)$ we can define the dual field $\tilde \pi(\tilde x)$ via the implicit relations\footnote{We choose a slightly different sign convention as in \cite{deRham:2013hsa} so that the fields are equivalent when $s=0$, $\tilde \pi(\tilde x)=\pi(x)+\mathcal{O}(s)$.} \cite{deRham:2013hsa}
\ba
\label{D1}
\D_s: \left\{\begin{array}{rcl}
x^\mu & \longrightarrow & \tilde x^{\mu} = x^{\mu} + \frac{s}{\Lambda^{\sigma}}\partial^{\mu} \pi(x) \,  ,\\[5pt]
\varphi_\mu(x)=\p_\mu \pi(x)& \longrightarrow & \tilde \varphi_\mu(\tilde x)=\tilde \p_\mu \tilde \pi(\tilde x)=\varphi_\mu(x)
\end{array}\right.\,.
\ea
Here $\Lambda$ is a fixed energy scale and $s$ is the parameter of the group transformation and $\sigma=d/2+1$ where $d$ is the number of spacetime dimensions. We can always choose $\Lambda$ such that $s=1$, however it is helpful for now to keep it distinct to make clear that there is a one parameter family of such transformations.

This transformation has a inverse, $\D_s^{-1}=\D_{-s}$,
\ba
\D_{-s}: \left\{\begin{array}{rcl}
\tilde x^\mu & \longrightarrow & x^{\mu} = \tilde x^{\mu} - \frac{s}{\Lambda^{\sigma}}\tilde{\partial}^{\mu} \tilde \pi(\tilde x) \,  ,\\[5pt]
\tilde \p_\mu \tilde \pi(\tilde x) & \longrightarrow & \p_\mu \pi(x)=\tilde \p_\mu \tilde \pi(\tilde x)
\end{array}\right.\,.
\ea
These implicit relations can equivalently be written as
\ba
\label{D1 pi}
\D_s:  \pi(x)&\longrightarrow & \tilde \pi(\tilde x) = \pi( x) + \frac{s}{2 \Lambda^{\sigma}} (\partial \pi(x))^2 \, ,\\
\D_{-s}:  \tilde \pi(\tilde x)&\longrightarrow & \pi(x) = \tilde \pi(\tilde x) -\frac{s}{2 \Lambda^{\sigma}} (\tilde \partial \tilde \pi(\tilde x))^2 \,.
\ea
The previous relations can be derived by recognizing that the above duality map can be understood as a Legendre transform \cite{Curtright:2012gx}.
These relations will be useful in determining the duality map at the level of the Lagrangians.

\subsection{\stu origin}

In the previous relations the derivative of $\pi$ transforms as a scalar\footnote{By scalar we mean here scalar under diffeomorphisms, for which the transformation law is $\tilde S(\tilde x)=S(x)$.} under the duality transformation
\be
\label{varphi}
\varphi_\mu(x)=\partial_{\mu} \pi(x) = \tilde \partial_{\mu} \tilde \pi (\tilde x)=\tilde \varphi_\mu(\tilde x) \, .
\ee
This result may at first seem surprising, however it is a natural consequence of the origin of the duality map in massive gravity and bi--gravity \cite{deRham:2010ik,deRham:2010kj,Hassan:2011zd,Fasiello:2013woa}.
There, the  `Galileon' $\pi(x)$ arises as the helicity-zero mode of a massive graviton. In the diffeomorphism invariant representation of massive gravity, there are four \stu fields $\phi^a$ which transform as scalars under diffeomorphisms and have an additional global Lorentz symmetry carried by the index $a$. It is always consistent to set the helicity-one mode to zero, and these \stu scalar fields can be expressed as $\phi^a = x^a +\frac{1}{\Lambda^{\sigma}} \partial^a \pi(x)$.

In the decoupling limit (when gravity is switched off) one can identify the global Lorentz symmetry with the global Lorentz subgroup of the local diffeomorphism symmetry. Under the combined group $\pi$ then transforms as a scalar. However beyond the decoupling limit the global Lorentz symmetry cannot be identified with the local diffeomorphisms, $\pi$ is not a scalar. Rather it is the derivative of $\pi$ (which is a vector under global Lorentz transformations) that is a scalar under the duality diffeomorphism. The transformation \eqref{varphi} is therefore the appropriate one for mode $\pi$. We prove this in detail in section~\ref{MGduality}.

\subsection{Duality Group}
A useful property of the duality map is that it forms a continuous group. To see this, let us perform a second duality transformation with parameter $s'$ starting from $\tilde \pi(\tilde x)$. Denoting the new dual field as $\hat \pi(\hat x)$ we then have by definition
\ba
%\D_s: \left\{\begin{array}{rcl}
%x^\mu & \to&  \tilde x^\mu = x^\mu +\frac{s}{\Lambda^\sigma}\p^\mu\pi (x)\\
%\pi(x) &\to& \tilde \pi (\tilde x)= \pi(x)+\frac{s}{\Lambda^\sigma}\(\p \pi(x)\)^2
%\end{array}
%\right.
%\\
\D_{s'}: \left\{\begin{array}{rcl}
\tilde x^\mu & \to&  \hat x^\mu = \tilde x^\mu +\frac{s'}{\Lambda^\sigma}\tilde \p^\mu\tilde \pi (x)= x^\mu+\frac{s+s'}{\Lambda^\sigma}\p^\mu\pi (x)\\
\tilde \pi(\tilde x) &\to& \hat \pi (\hat x)= \tilde \pi(\tilde x)+\frac{s'}{\Lambda^\sigma}\(\tilde \p \tilde \pi(\tilde x)\)^2
=\pi(x)+\frac{s+s'}{\Lambda^\sigma}\(\p \pi(x)\)^2
\end{array}
\right. \,,
\ea
where we used the relations \eqref{D1} and \eqref{D1 pi}. This leads to  a combined transformation
\ba
\D_{s'}\circ\D_s = \D_{s+s'}\,.
\ea

%\ba
%&& \hat x^{\mu} = \tilde x^{\mu} + \frac{s'}{\Lambda^{\sigma}}\tilde \partial^{\mu} \tilde \pi(\tilde x) \,  ,\\
%&& \tilde x^{\mu} = \hat x^{\mu} - \frac{s'}{\Lambda^{\sigma}}\hat \partial^{\mu} \hat \pi(\hat x) \, .
%\ea
%and
%\ba
%&& \tilde \pi(\tilde x) = \hat \pi(\hat x) -\frac{s'}{2 \Lambda^{\sigma}} (\hat \partial \hat \pi(\hat x))^2 \,, \\
%&& \hat \pi(\hat x) = \tilde \pi( \tilde x) + \frac{s'}{2 \Lambda^{\sigma}} (\tilde \partial \tilde \pi(\tilde x))^2 \, .
%\ea
%Combining these relations with the previous ones it is easy to show that
%\ba
%&& \hat x^{\mu} = x^{\mu} + \frac{s+s'}{\Lambda^{\sigma}}\partial^{\mu} \pi(x) \,  ,\\
%&&  x^{\mu} = \hat x^{\mu} - \frac{s+s'}{\Lambda^{\sigma}}\partial^{\mu} \hat \pi(\hat x) \, .
%\ea
%and
%\ba
%&& \pi(x) = \hat \pi(\hat x) -\frac{s+s'}{2 \Lambda^{\sigma}} (\hat \partial \hat \pi(\hat x))^2 \,, \\
%&& \hat \pi(\hat x) = \pi( x) + \frac{s+s'}{2 \Lambda^{\sigma}} (\partial \pi( x))^2 \, .
%\ea
In other words the duality map forms an abelian group with transformation law $s''=s+s'$. The inverse group transformation corresponds to $s' = -s$.
Again we note that this group transformation leaves invariant the derivatives of the Galileon fields
\ba
\partial_{\mu} \pi(x) = \tilde \partial_{\mu} \tilde \pi (\tilde x) = \hat \partial_{\mu} \hat \pi (\hat x)\, .
\ea
In the Galileon theories and in massive gravity/bi--gravity, particular importance is placed in the Galileon invariant combination
 \ba
\Pi\mupn(x)=\frac{1}{\Lambda^{\sigma}}\eta^{\mu\alpha}\partial_\alpha\partial_{\nu} \pi(x)\,,
\ea
and similarly $\tilde \Pi\mupn(x)=\tilde \partial^{\mu}\tilde \partial_{\nu} \tilde \pi(\tilde x)/\Lambda^{\sigma}$ and with the hat variables. In terms of these quantities we then have (suppressing indices and using matrix notation)
\ba
\label{Pi 1}
&& \tilde \Pi = \left[\mathbb{I} +s \Pi \right]^{-1} \Pi  \, ,\\
\label{Pi 2}
&& \hat \Pi = \left[ \mathbb{I} + s' \tilde \Pi \right]^{-1} \tilde \Pi  \, , \\
\label{Pi 3}
&& \hat \Pi = \left[ \mathbb{I} + (s+s') \Pi \right]^{-1} \Pi \, .
\ea
Here it is understood that $\Pi$ is evaluated at $x$, $\tilde \Pi$ at $\tilde x$ and $\hat \Pi$ at $\hat x$.
These relations are equivalently written as
\ba
\label{Pi 4}
&& \tilde \Pi^{-1} = \Pi^{-1}+s\  \mathbb{I}  \, ,\\
\label{Pi 5}
 && \hat \Pi^{-1} = \tilde \Pi^{-1}+ s'\   \mathbb{I}\, , \\
\label{Pi 6}
 && \hat \Pi^{-1} = \Pi^{-1}+ (s+s')\  \mathbb{I} \, .
\ea
In this latter form the abelian group property is manifest. This means that a finite duality transformation may be built out of infinitesimal ones for which the infinitesimal variation is
\be
\delta \pi(x) = \tilde \pi(x) - \pi(x)= - \frac{s}{2 \Lambda^\sigma}  (\partial \pi(x))^2+ {\cal O}(s^2) \, .
\ee
We see that the infinitesimal transformation is a {\bf local} field redefinition. This result is one way to understand why the duality preserves the notion of locality. The naively non-local finite $s$ duality map can be viewed as an infinite number of local infinitesimal transformations. This is analogous to viewing a large gauge transformation as an infinite number of infinitesimal ones.
Since it is known that the S-matrix is invariant under perturbative local invertible field redefinitions, it is then invariant under the infinitesimal transformation, \ie
\be
\delta \langle f |  \hat S | i \rangle=\frac{\d \langle f |  \hat S | i \rangle}{\d s} \delta s =0 \, .
\ee
But the continuous group property now implies that this can be integrated to finite $s$ confirming that the S-matrix is {\it invariant under the full duality map}.

\subsection{Vector Field Duality}

As we have already discussed the scalar field duality has a natural interpretation in the context of massive gravity and bi--gravity theories \cite{deRham:2010ik,deRham:2010kj,Hassan:2011zd,Fasiello:2013woa} as different choices of gauges for the \stu fields. In general as well as a scalar component representing the helicity-zero mode of the massive graviton, there is also a vector component representing the helicity-one mode of the massive gravity. More generally the \stu fields can be written as follows
\be
\phi^a = x^a + \frac{1}{\Lambda^{\sigma-1}}A^a(x) \,.
\ee
This prompts the definition of a second group of transformations which depends entirely on the vector fields
\ba
\label{D2}
\mathbb{D}_t: \left\{\begin{array}{rcl}
x^\mu & \longrightarrow & \tilde x^{\mu} = x^{\mu} + \frac{t}{\Lambda^{\sigma-1}}A^{\mu}(x) \,  ,\\[5pt]
A_{\mu}(x)& \longrightarrow & \tilde A_{\mu}(\tilde x)= A_{\mu}(x)
\end{array}\right.\,,
\ea
and has the inverse
\ba
\label{D2 inverse}
\mathbb{D}_{-t}: \left\{\begin{array}{rcl}
\tilde x^\mu & \longrightarrow &  x^{\mu} = \tilde x^{\mu} - \frac{t}{\Lambda^{\sigma-1}}\tilde A^{\mu}(\tilde x) \,  ,\\[5pt]
\tilde A_{\mu}(\tilde x)& \longrightarrow &  A_{\mu}( x)= \tilde A_{\mu}(\tilde x)
\end{array}\right.\,.
\ea
As in the scalar case these transformations form an abelian group
\ba
\mathbb{D}_{t'}\circ\mathbb{D}_t = \mathbb{D}_{t+t'}\,,
\ea
since
\ba
%\D_s: \left\{\begin{array}{rcl}
%x^\mu & \to&  \tilde x^\mu = x^\mu +\frac{s}{\Lambda^\sigma}\p^\mu\pi (x)\\
%\pi(x) &\to& \tilde \pi (\tilde x)= \pi(x)+\frac{s}{\Lambda^\sigma}\(\p \pi(x)\)^2
%\end{array}
%\right.
%\\
\mathbb{D}_{t'}\circ \mathbb{D}_t:
x^\mu  \to  \hat x^\mu = \tilde x^\mu +\frac{t'}{\Lambda^{\sigma-1}}\tilde A_{\mu} (\tilde x)= x^\mu+\frac{t+t'}{\Lambda^{\sigma-1}} A_{\mu} (x)\,,
\ea
and we have
\ba
\hat A_{\mu}(\hat x) = \tilde A_{\mu}(\tilde x) = A_{\mu}(x)\,.
\ea
Thus again we see that $A_{\mu}$ transforms as a diffeomorphism scalar under the duality transformation despite being a global Lorentz vector.

The equivalent of the relations (\ref{Pi 1}-\ref{Pi 6}) is true for $B_\mu^{\ \nu}=\p_{\mu}A^{\nu}/\Lambda^{\sigma-1}$ with
\ba
\tilde B_\mu^{\ \nu} (\tilde x)=\(\delta_\mu^{\ \alpha}-t \tilde B_\mu^{\ \alpha}(\tilde x)\)B_\alpha^{\ \nu}(x)\,,
\ea
or introducing matrix notation
\be
\(\mathbb{I}+tB(x)\) = \(\mathbb{I}-t \tilde B(\tilde x)\)^{-1}\,.
\ee

\section{Duality for Generalized Galileons}
\label{sec:GeneralizedGalileon}

\subsection{Generalized Galileons}

In recent years there has been a revival of interest in writing down the most general expression for the Lagrangian for scalar field theories that have second order equations of motion. Restricting ourselves to a single scalar field on Minkowski spacetime, the most general local and Lorentz invariant Lagrangian which do not suffer from the Ostrogradsky instability is the generalization of the Galileon that takes the form \cite{Deffayet:2011gz}
\be\label{gg}
S=\int \d^dx\sum_{n=0}^{d} A_n(\pi, X)\,  \, {\cal U}_n[\Pi(x)]\,,
\ee
where $X= -\frac{1}{2} (\partial \pi)^2$. This is the Horndeski Lagrangian restricted to Minkowski spacetime.
The Lagrangian includes within it $k$-essence \cite{ArmendarizPicon:2000dh,ArmendarizPicon:2000ah}, Galileons and canonical scalar fields with potentials as special cases.
Here $A_n(\pi,X)$ are arbitrary functions of $\pi$ and $X$.
For any matrix $\mathbb{X}$ we have defined the usual characteristic polynomials
\ba
\, {\cal U}_n[\mathbb{X}]=\ep^{\mu_1 \cdots \mu_d}\ep^{\nu_1 \cdots \nu_d}
\prod_{j=1}^n \mathbb{X}_{\mu_j\nu_j}\prod_{k=n+1}^d \eta_{\mu_k\nu_k}
\,,
\ea
where $\ep$ is the Levi-Civita symbol. In what follows we use the notation $1=\mathbb{I}$ when no confusion can arise. An equivalent definition is through the determinant
\be
{\rm det}[1+\lambda \mathbb{X}] = \sum_{n=0}^d \frac{1}{n! (d-n)!} \lambda^n \, {\cal U}_n[\mathbb{X}]\,.
\ee

The Galileon \cite{Nicolis:2008in} corresponds to the case where the $A_n(\pi,X)$ take the form $A_n(\pi,X) = c_n \pi$, with constant coefficients $c_n$.  This form preserves the Galileon symmetry $\pi \rightarrow \pi + v_{\mu} x^{\mu}$ at the level of the action and equations of motion, but not at the level of the Lagrangian.

Let us now consider how this action changes under a duality transformation with parameter $s$.
For this we need to utilize the Jacobian of the transformation from $x$ to $\tilde x$
\ba
\left|\frac{\partial  x^a}{\partial \tilde x^b}\right|=\det\(1+s \Pi(x)\)^{-1} =\det\(1-s \Sigma(\tilde x)\) \, ,
\ea
assuming that the sign of the determinant is positive\footnote{If the determinant changes sign it must pass through zero at which point the field redefinition is not technically invertible. For this reason we confine our attention to the branch for which the sign is always positive. This is similar to GR where the determinant of the metric does not pass through zero within the regime of validity of that theory.\label{footnote:det}} and from now on  we use the notation $\tilde \pi \equiv \rho$ to make the different representations more manifest and write $\Sigma\mupn(\tilde x)= \tilde \p^\mu \tilde \p_\nu \rho(\tilde x)/\Lambda^\sigma$, with
\ba
\Sigma=\left[\mathbb{I} +s \Pi \right]^{-1} \Pi \,.
\ea

Setting $\Lambda=1$ (or absorbing it in $s$),
and using the relations $\pi(x) = \rho(\tilde x) -\frac{s}{2} (\tilde \partial \rho(\tilde x))^2$, $X=-1/2 (\partial \pi)^2= -1/2 (\tilde \partial \rho(\tilde x))^2$ and $ \Sigma = \left[1 +s \Pi \right]^{-1} \Pi$, into the Lagrangian we soon find
\be
S_s=\int \d^d \tilde x \det\(1-s \Sigma(\tilde x)\) \, \sum_{n=0}^{d}A_n\(\rho(\tilde x) +s X, X\) \, {\cal U}_n\left[\frac{\Sigma(\tilde x)}{1-s \Sigma(\tilde x)}\right]\,.
\ee
Now since $\tilde x$ is a dummy integration variable we can equivalently write this as
\be
S_s=\int \d^d  x \det\(1-s \Sigma(x)\) \, \sum_{n=0}^{d}A_n\(\rho(x) +s Y, Y\) \, {\cal U}_n\left[\frac{\Sigma(x)}{1-s \Sigma( x)}\right]\,,
\ee
where to avoid confusion we have now defined $Y=-1/2(\partial \rho(x))^2$. Finally after a straightforward rearrangement we have the final form for the dual Lagrangian
\be
S_s=\int \d^dx\sum_{n=0}^{d}B_{n, s}(\rho, Y) \, {\cal U}_n[\Sigma(x)]\,,
\ee
where the functions $B_{n, s}$ are linear combinations of the original functions in the form
\ba
\label{coefficients}
B_{n, s}(\rho, Y)=\sum_{k=0}^{d}(-1)^{n-k} s^{n-k} A_{k}\(\rho +s Y, Y \) \frac{(d-k)!}{(n-k)!(d-n)!}\,,
\ea
with $n!= \Gamma(n+1)$. We thus see that every generalized Galileon is dual to a one parameter family of other generalized Galileons.

\subsection{Dual of a Canonical Scalar}

As an illustrative example, let us consider the following case of a canonical scalar field with a potential
\be
S = \int \d^dx \, \( X-V(\pi) \)\,.
\ee
Following the recipe this is dual to a generalized version of the quintic (in 4 dimensions) or $(d+1)^{\rm th}$ order Galileon considered in \cite{deRham:2013hsa}. Explicitly this takes the form
\be
S_s = \int \d^dx \, \det\(1-s \Sigma(x)\) \left[ Y-V(\rho+s Y) \right]\,.
\ee
As an extension of the result of \cite{deRham:2013hsa} (see also \cite{Creminelli:2013fxa}), a free massive scalar field with $V(\pi) = \frac{1}{2} m^2 \pi^2$ is dual to the following
\be\label{free}
S_s = \int \d^dx \, \det\(1-s \Sigma(x)\) \left[ -\frac{1}{2}(\partial \rho)^2-\frac{1}{2} m^2 \(\rho-\frac{s}{ 2} (\partial \rho)^2\)^2 \right]\, .
\ee
This means that this theory is a free theory, regardless of the value of $s$, \ie all its tree level scattering amplitudes vanish. By using the optical theorem then we infer that the loops have no imaginary parts and so we may then argue that there is an appropriate way to quantize the theory (\ie appropriate choice of path integral measure) where the scattering amplitudes vanish to all orders. This also implies that this local, naively non-renormalizable Lagrangian with strong coupling scale $\Lambda$ is actually UV complete by itself.

\subsection{Duality as a Symmetry}

As we have emphasized earlier, the duality transformation forms an abelian group. However in general the action is not invariant under this transformation only the S-matrix is (since the S-matrix is invariant under field redefinitions). However we will now show that there is a special choice for which the duality becomes a true symmetry. Since the group is continuous, we consider an infinitesimal transformation $s$ for which
\be
B_{n, s}(\pi,X) = A_n(\pi,X) - s (d-n+1) A_{n-1}(\pi,X) + s X \frac{\partial}{\partial \pi} A_n(\pi,X) + {\cal O}(s^2)\,.
\ee
We thus infer that the action is left invariant under an infinitesimal duality transformation provided that the coefficient functions satisfy the recursion relation
\be
A_{r}(\pi,X) = \frac{X}{(d-r)} \frac{\partial}{\partial \pi} A_{r+1}(\pi,X)\,, \hspace{20pt}\forall\hspace{5pt}0\le r < d-1\, .
\ee
This can be solved to give
\be
\label{Sym relations}
A_r(\pi,X) = \frac{X^{d-r}}{(d-r)!} \frac{\partial^{(d-r)}}{\partial \pi^{(d-r)}} A_d(\pi,X) \, .
\ee
Hence there is an infinite family of actions for which the duality transformation becomes a symmetry, specified only by $A_d(\pi,X)$.

\subsubsection*{$\bullet$ Example of  Symmetric theories}

As a simple example of this we note that any function of the form $A_d(\pi, X) =\pi F(X)$ will lead to the simple action
\ba
\label{SF}
S_{F}=\int \d^d x \frac{1}{\Lambda^{(d-1)\sigma}}F(X)\(-\frac12 (\p \pi)^2 \, {\cal U}_{d-1}[\Pi]+\pi \, {\cal U}_{d}[\Pi]\)\,,
\ea
where we included the scale $\Lambda$ for consistency and so $X=-\frac1{2\Lambda^d} (\p \pi)^2$. This is one simple example of duality invariant theory (up to a total derivative),
\ba
\D_s: \hspace{10pt} S_{\rm F} \hspace{5pt} \longrightarrow \hspace{5pt} S_{\rm F}\,.
\ea
If $F(X)$ was chosen to be a constant $F(X)=F_0$ we recover a special Galileon (a quintic Galileon in four dimensions, the highest possible Galileon in arbitrary dimensions),
\ba
\label{SF}
S_{\rm F_0}&=&\int \d^d x \frac{F_0}{\Lambda^{(d-1)\sigma}}\(-\frac12 (\p \pi)^2 \, {\cal U}_{d-1}[\Pi]+\pi \, {\cal U}_{d}[\Pi]\)\\
&=&\frac{d+2}{d+1} \int \d^d x \frac{F_0}{\Lambda^{(d-1)\sigma}}\pi \, {\cal U}_{d}[\Pi]\,.
\ea
This is consistent with the results presented in \cite{deRham:2013hsa} where it is clear that the highest order Galileon dualizes to itself.

This result is of course generalizable to $F(X)$ not constant where the Galileon symmetry is broken. This class of theories is generalizable to
\ba
\label{Sym eg}
S_{G}=\int \d^d x \frac{1}{\Lambda^{(d-2)\sigma}}\tilde G(X) \, {\cal U}_{d-1}[\Pi] = \int \d^d x \frac{1}{\Lambda^{(d-1)\sigma}}\pi G(X) \, {\cal U}_{d}[\Pi]\,,
\ea
where $G$ is non-trivially related to  $\tilde G$.  If $\tilde G$ is a polynomial of rank $r$ then $G$ is a polynomial of rank $r-1$.

The dual version of this action is
\ba
\D_s S_{G}[\pi,X,\Pi]=\int \d^d x \frac{1}{\Lambda^{(d-1)\sigma}}\det\(1-s \Sigma\) \(\rho+s Y\) G(Y) \, {\cal U}_{d}\left[\frac{\Sigma}{1-s \Sigma}\right]\,,
\ea
since
\ba
\det\(1-s \Sigma\)&=&1-s [\Sigma]+\mathcal{O}(s^2)\\
\, {\cal U}_{d}\left[\frac{\Sigma}{1-s \Sigma}\right]&=&\, {\cal U}_{d}[\Sigma]\(1+s[\Sigma]\)+\mathcal{O}(s^2)\,,
\ea
and since any Lagrangian of the form $L(Y)\, {\cal U}_{d}[\Sigma]$ is a total derivative for arbitrary function $L(Y)$,
we infer straight away that
\ba
\D_s S_{G}[\pi,X,\Pi]&=&S_G[\rho,Y,\Sigma]+ s  \int\d^d x Y G(Y)\, {\cal U}_{d}[\Sigma]+\mathcal{O}(s^2)\nn\\
&\equiv& S_G[\rho,Y,\Sigma]+\mathcal{O}(s^2) \,.
\ea
The theory \eqref{Sym eg} is invariant under the duality infinitesimal transformation.\\

As a more general class of examples we could consider any function of the form $A_{d, k}(\pi, X) =\pi^k F(X)$ for any positive power $k$. For $k=0$ the resulting Lagrangian is  a total derivative but for $k\ge 1$ we get a non-trivial class of theories for which the duality transformation is a global symmetry. This global symmetry could then be gauged to lead to a local symmetry. This would be an interesting avenue to explore.\\

These symmetric theories do not have a kinetic term when $\langle \pi\rangle=0$. However they do have a well-defined kinetic term as well as a tadpole after a Lorentz invariant shift of $\pi \to \check{\pi}=\pi -1/2 x^2\Lambda^\sigma$. After such a shift the Galileon duality takes the form of a Legendre transform \cite{Curtright:2012gx,deRham:2013hsa}
\ba
\check{\D}_s: x^\mu \to \tilde x^\mu= \frac{s}{\Lambda^\sigma}\p^\mu \check{\pi}\,,
\ea
and so all the theories presented in this section have a well-defined kinetic term for $\check{\pi}$ and are invariant under the transformation $\check{D}_s$.

\section{Duality for Vector Fields}

\label{sec:VectorDuality}

In this section we will consider the vector duality for some simple examples of vector theories.
In this case the Jacobian for the transformation is given by
\ba
\left|\frac{\partial  x^a}{\partial \tilde x^b}\right|=\det\(1+t B(x)\)^{-1} =\det\(1-t \tilde B(\tilde x)\) \, ,
\ea
again assuming that we restrict ourselves to the region for which the determinant is positive and so the transformation is always invertible (see footnote \ref{footnote:det}). Similarly we will make use of the relation
\be
B(x)= \frac{\tilde B(\tilde x)}{\mathbb{I} -t \tilde B(\tilde x)}\,,
\ee
remembering that $B_{\mu}{}^{\nu}(x)= \partial_{\mu} A^{\nu}(x)$ and similarly $\tilde B_{\mu}{}^{\nu}(\tilde x)= \tilde \partial_{\mu} \tilde A^{\nu}(\tilde x)$.

\subsection{Example of Massless spin-1: Dual of Maxwell's theory}

To begin with let us consider the Maxwell action in $d$-dimensions
\be
\label{Maxwell1}
S_{\rm Maxwell} = \int \d^d x \(- \frac{1}{4} F_{\mu\nu} F^{\mu\nu} \)\, ,
\ee
where $F_{\mu\nu} = \partial_{\mu} A_{\nu}- \partial_{\nu} A_{\mu}$. This Lagrangian is manifestly gauge invariant. However it may be equivalent written in the form
\be
S_{\rm Maxwell} = \int \d^d x\( - \frac{1}{2} \left( \partial_{\mu}A_{\nu} \partial^{\mu}A^{\nu}-  \partial_{\mu}A_{\nu} \partial^{\nu}A^{\mu} \right)\) \, ,
\ee
in which gauge invariance is not manifest at the level of the Lagrangian. In matrix language this is the statement that
\be
S_{\rm Maxwell} = \int \d^d x \(- \frac{1}{2} \left( {\rm Tr}[B^T B] - {\rm Tr}[B^2] \right)\)\, .
\ee
Now following the previous steps this action is dual to the following
\ba
\label{dual Maxwell}
\tilde S_{{\rm Maxwell}, t} = \int \d^d x  \(- \frac{1}{2} {\rm det}(1-t \tilde B)\left( {\rm Tr}\left[\(\frac{\tilde B}{1-t \tilde B}\)^T \frac{\tilde B}{1- t \tilde B}\right] - {\rm Tr}\left[ \frac{\tilde B^2}{(1- t \tilde B)^2}\right] \right)\)\, . \nn
\ea

\subsubsection*{$\bullet$ Dual of the $U(1)$--symmetry}

This is a remarkable feature. This theory has a $U(1)$-gauge invariance since it is dual to Maxwell theory and only propagates two degrees of freedom yet  it is not built out of the gauge invariant quantity $\tilde F\mn$ nor $^* \tilde{F}\mn$. This comes to show that a $U(1)$-gauge invariant theory can take a very different form involving terms which are not expressible in terms of $\tilde F\mn$. The reason for that is that the realization of the $U(1)$ symmetry is in this case non-linear in the field and very non-trivial. To derive its explicit form we start with the $U(1)$ in the original $\pi$-duality frame where
\ba
A_\mu(x) \xrightarrow{U(1)}A'_\mu(x)=A_\mu(x)+\p_\mu \theta(x)\,.
\ea
Now recalling that the duality transformation acts as follows
\ba
A_\mu(x)\xrightarrow{\ \mathbb{D}_t\ } \tilde{A}_\mu(\tilde x)=A_\mu(x)\hspace{10pt}{\rm with}\hspace{10pt}
\tilde x^\mu=x^\mu+t  A^\mu(x)\,,
\ea
we deduce that $\tilde A \xrightarrow{U(1)}\tilde A' $ with
\ba
\tilde A'_\mu(x+t A' )=\tilde A'_\mu (\tilde x^\nu +t \p^\nu \theta(x)) = A_\mu(x)+\p_\mu \theta(x) \,.
\ea
We thus infer that
\ba
\tilde A_\mu(\tilde x)  \xrightarrow{U(1)}  \tilde A'_\mu(\tilde x)&=&\tilde A_\mu(\tilde x)+\(\delta^\nu_\mu- t \tilde \p^\nu \tilde A_\mu(\tilde x) \) \p_\nu \theta(x)\nn \\
\label{U1 dual}
&=&\tilde A_\mu(\tilde x)+\Big(\(1-t\tilde B\)^T\Big)^\nu_{\ \, \mu}  \Big(\(1-t \tilde B\)^{-1}\Big)^{\ \,  \alpha}_{\nu} \tilde \p_\alpha \tilde \theta(\tilde x)\,.
\ea
This is a highly non-linear representation of a $U(1)$ transformation. Nevertheless it must form an abelian group and satisfy the same properties as a $U(1)$. One can check explicitly that the Lagrangian \eqref{dual Maxwell} is invariant under this symmetry. We shall show it at leading order in $t$ below.

\subsubsection*{$\bullet$ Leading order in $t$}

At leading order in the transformation parameter $t$,  the theory \eqref{dual Maxwell} is
\ba
\label{Maxwwell t 1}
\tilde S_{{\rm Maxwell}, t} = \int \d^d x  \(
-\frac 14 \tilde{F}\mn^2+ t \left[\frac 14 [\tilde{B}]\tilde{F}^2\mn +[\tilde{B}^2 \tilde{F}] \right]+\mathcal{O}(t^2)
\)\,.
\ea
Already at this level the theory is not expressible solely in terms of $\tilde{F}\mn$. Yet it is invariant under the non-linearly realized $U(1)$-transformation \eqref{U1 dual} at leading order in $t$ given by (after relabeling the dummy variable $\tilde x$ to $x$ as usual),
\ba
\tilde B'\mn&=&\tilde B\mn+\p_\mu \p_\nu \tilde \theta+t \p_\mu\(\tilde F_{\nu}^{\  \alpha}\p_\alpha \tilde \theta\)+\mathcal{O}(t^2)\\
\tilde F'\mn&=& \tilde F\mn+ t \left[2\tilde F_{[\nu}^{\ \alpha}\p_{\mu]} \p_\alpha \tilde \theta-\p^\alpha \tilde F\mn \p_\alpha \tilde\theta\right]
+\mathcal{O}(t^2) \,.
\ea
Up to that order, the transformation of the first term in \eqref{Maxwwell t 1} is then
\ba
\delta_{\tilde \theta}\left[-\frac 14 \tilde F\mn^2\right] &=& - t \left[ \tilde F^{\mu\alpha} \tilde F_{\alpha}^{\ \nu}\p_\mu \p_\nu \theta -\frac 12 \tilde F^{\mu\nu}\p^\alpha \tilde F\mn \p_\alpha \tilde \theta \right] \nn\\
&=& - t \left[ \tilde F^{\mu\alpha} \tilde F_{\alpha}^{\ \nu}\p_\mu \p_\nu \theta + \frac 14  \tilde F\mn^2 \Box \tilde \theta \right]\,,
\ea
and the second term transforms as
\ba
t \delta_{\tilde \theta} \left[\frac 14 [\tilde{B}]\tilde{F}^2\mn +[\tilde{B}^2 \tilde{F}]\right]
 =  t \left[\frac 14 \Box \tilde \theta \tilde F\mn^2 +\tilde F^{\mu\alpha}\tilde F_\alpha^{\ \nu}\p_\mu \p_\nu\tilde \theta \right]\,,
\ea
so the dual to the Maxwell Lagrangian is clearly invariant under the transformation \eqref{U1 dual} to leading order in $t$. The fact that \eqref{dual Maxwell} should remain invariant under \eqref{U1 dual} to all orders in $t$ is of course a simple consequence to the $U(1)$ symmetry in Maxwell theory. So we emphasize once more that \eqref{dual Maxwell} propagates only two degrees of freedom in four dimensions and enjoys a $U(1)$ gauge symmetry,  yet \eqref{dual Maxwell} is not expressible in terms of only $F\mn$.

This could serve as an inspiration when building the most general $U(1)$-gauge invariant theory as it allows for the possibility of new terms which were not considered before. This could potentially lead to a generalization of \cite{Deffayet:2013tca}.

\subsection{Example of Massive spin-1: Dual of Proca's theory}

As another simple example, let us now consider a Proca theory
\ba
\label{Proca1}
S_{\rm Proca} = \int \d^d x \(- \frac{1}{4} F_{\mu\nu} F^{\mu\nu} -\frac 12 m^2 A_\mu^2\)\, ,
\ea
which propagates $d-1$ degrees of freedom. Using the same derivations as before, we infer that the dual to this Proca's theory is
\ba
\label{Proca2}
\tilde{S}_{{\rm Proca}, t} = \tilde{S}_{ {\rm Maxwell }, t}-\frac12 m^2 \int \d^d x \det\(1-t \tilde B\)\tilde A^2_\mu\, ,
\ea
where $\tilde{S}_{\rm Maxwell }$ is given in \eqref{dual Maxwell}. It would be interesting to see if this lies within the class of theories explored in \cite{Tasinato:2014eka,Heisenberg:2014rta} or if they correspond to a new class of interactions which still propagate the correct number of degrees of freedom.

\section{Coupling to Matter}

\label{mattercoupling}

\subsection{Arbitrary Fields}

We now include arbitrary coupling of the Generalized Galileon field to matter. This is  remarkably straightforward to do. The key is to recognize that the duality is itself just a specific field dependent diffeomorphism. As already explained $\pi$ is not a scalar under diffeomorphisms, but its derivative is.  Matter on the other hand should transform as it does normally under a diffeomorphism. For instance for a scalar field $\chi(x)$ we define the dual scalar $\tilde \chi$ via the relation
\be
\label{mattertransform}
\D_s: \chi(x)\to \tilde \chi(\tilde x) = \chi(x)\,,
\ee
and where we continue to use $ \tilde x^{\mu} = x^{\mu} + \frac{s}{\Lambda^{\sigma}}\partial^{\mu} \pi(x) $. For a vector field the dual vector is defined as
\be
\D_s: V_{\mu}(x)\to \tilde V_{\mu}(\tilde x)= \frac{\delta x^\nu}{\delta \tilde x^\mu }V_\nu(x) =
 \left[ 1-s \Sigma(\tilde x) \right]_{\ \mu}^{\nu}  V_{\nu}(x)\,,
\ee
which may be equivalently written as
\be
V_{\mu}(x) =  \left[ 1+ s \Pi(x) \right]_{\ \mu}^{\nu} \tilde V_{\nu}(\tilde x) \, .
\ee
From this it is straightforward to generalize to an arbitrary tensor field
\be
\D_s: T_{\mu_1 \dots \mu_r}(x) \to \tilde T_{\mu_1  \dots \mu_r}(\tilde x) =
 \left[ 1-s \Sigma(\tilde x) \right]_{\ \mu_1}^{\nu_1} \cdots\left[ 1-s \Sigma(\tilde x) \right]_{\ \mu_r}^{\nu_r}T_{\nu_1 \dots \nu_r}(x) \,.
%T_{\mu_1 \mu_2 \mu_3 \dots}(x) = \tilde T_{\nu_1, \nu_2, \nu_3 \dots}(\tilde x)  \left[ 1+ s \Pi(x) \right]_{\mu_1}^{\nu_1}\left[ 1+ s \Pi(x) \right]_{\mu_2}^{\nu_2}\left[ 1+ s \Pi(x) \right]_{\mu_3}^{\nu_3} \dots \, ,
\ee
Fermions should be viewed as if they are living on a curved spacetime. Thus for example in a curved spacetime Dirac spinors $\Psi_{\alpha}$ are representations of the local Lorentz group, and thus transform only as scalars under diffeomorphisms
\be
\D_S: \Psi_{\alpha}(x)\to \tilde \Psi_{\alpha}(\tilde x) = \Psi_{\alpha}(x) \, .
\ee

\subsection{General Matter Lagrangian}

Let us now consider an arbitrary action for the matter, assumed for simplicity to be a scalar $\chi$, including coupling to the Generalized Galileon field $\pi$ of the form
\be
\label{eq3}
S_{\rm matter} = \int \d^d x \, {\cal L}_{\rm matter} ( \chi(x), \partial_{\mu} \chi(x) , \pi(x), \partial_{\mu} \pi(x)) \, ,
\ee
where it is understood that $\partial_{\mu} \chi(x)$ and $\partial_{\mu} \pi(x)$ are built into scalar combinations using either the Minkowski metric or the Levi-Civita symbols.
This form certainly covers the types of couplings expected for almost all well-defined theories.

Following our previous recipe with the above transformation rules for the matter, the dual matter Lagrangian takes the form
\ba
\label{eq2}
\tilde S_{{\rm matter}, s} = \int \d^d x \,  \det\(1-s \Sigma\)  {\cal L}_{{\rm matter}} \( \tilde \chi, [(1-s \Sigma)^{-1} ]_{\ \mu}^{\nu}\partial_{\nu} \tilde \chi , \rho- \frac{s}{2\Lambda^{\sigma} }(\partial \rho)^2, \partial_{\mu} \rho\)\! , \ \
\ea
where in the previous expression $\tilde \chi$ and $\rho$ are evaluated at $x$.

Now the key point is that this Lagrangian is manifestly local. It clearly remains local for $\tilde \chi$ since it depends on no more that $\tilde \chi(x)$ and $\partial_{\mu} \tilde \chi(x)$. It is also clearly local in $\rho$ since it only depends on $\rho(x)$, $\partial_{\mu} \rho(x)$ and $\partial_{\mu}\partial_{\nu} \rho(x)$. It does however contain one apparent wrinkle. The dependence of the matter Lagrangian on $\partial_{\mu}\partial_{\nu} \rho(x)$ would appear to suggest the presence of an Ostrogradsky ghost \cite{Ostrogradsky}. For instance if we find the equation of motion for $\tilde \chi$ it would certainly contain in general triple derivatives of $\rho$.

However this is an example of a phenomena that was observed in \cite{deRham:2011rn,deRham:2011qq}. It is possible that the equations of motion for a dynamical system with multiple fields derived directly from the Lagrangian do contain higher than two derivatives without this implying new degrees of freedom provided that it is possible to rearrange the equations of motion and their derivatives in a form for which the higher derivatives cancel. In other words, suppose we have the equations of motion ${\cal E}_{\rho}=0$ for $\rho$ and ${\cal E}_{\tilde \chi}=0$ for $\tilde \chi$. Provided that there is a combination of the equations of motion of the form
\ba
{\cal E}'_{\rho}={\cal E}_{\rho}+ C_1 \frac{d}{dt}{\cal E}_{\tilde \chi}+C_2 \,  {\cal E}_{\tilde \chi} \\
{\cal E}'_{\tilde \chi}={\cal E}_{\tilde \chi}+ D_1 \frac{d}{dt}{\cal E}_{\rho} +D_2\,  {\cal E}_{\rho}
\ea
for which the two ${\cal E}'$ do not contain higher than two time derivatives of either $\rho$ or $\tilde \chi$, then the equations of motion remain second order.

This must be the case here since the Galileon duality map is an invertible transformation and so cannot change the number of propagating degrees of freedom.

 \subsubsection{Specific Example}
 \label{sec:example}

 To see that this is the case it is helpful to work with a simple example. Consider the case of a specific Galileon coupled to a canonical scalar matter field through a simple $\chi^2 \pi$ coupling.
In the $\pi$-representation, we consider the following  action\footnote{The specific example \eqref{example} is chosen for pedagogy and definiteness but none of the arguments are specific to this case.}
\ba
\label{example}
S = \int \d^d x  \left[ -{\rm det}(1+ \Pi) \frac{1}{2}(\partial \pi)^2- \frac{1}{2} (\partial \chi)^2 + g \chi^2 \pi \right] \, .
\ea
In the rest of this paper, this action will continue to serve as an archetype of a Galileon coupled to a scalar field, which is the reason why the $\det(1+ \Pi)$  is introduced. In section \ref{sec:SL} we will see how about for generic backgrounds the kinetic structure $\det(1+ \Pi)(\partial \pi)^2$ leads to superluminal group velocity at the classical level. However as is well known, classical superluminal group velocity do not imply acausalities as shall be discussed in section \ref{sec:SL}.

\subsubsection*{$\bullet\ \pi(x)$-duality frame}

In the $\pi$-duality frame, the equations of motion take the form
\be
\label{E_chi}
\mathcal{E}_\chi=\Box \chi +2 g \pi \chi=0 \, ,
\ee
and
\be
\label{E_pi}
\mathcal{E}_\pi={\rm det}[1+\Pi] {\rm Tr}\left[\frac{\Pi}{1+\Pi}\right] + g \chi^2 =0 \, .
\ee
In this form it is clear that the equations of motion are second order in derivatives which is just the well known result for the Galileon.

\subsubsection*{$\bullet\ \rho(\tilde x)$-duality frame}

Now let us perform the duality transformation with $s=1$. The dual action is then
\ba
\tilde{S}_s&=&\int \d ^d x \Bigg(-\frac 12 (\p \rho)^2+ g \det\(1-\Sigma\) \tilde \chi^2 \(\rho-\frac 1{2 \Lambda^{\sigma}} (\p \rho)^2\)\\
&&-\frac 12 \det\(1-\Sigma\)[\(1-\Sigma\)^{-1}]^\mu_{\ \alpha}[\(1-\Sigma\)^{-1}]^\nu_{\ \beta}\eta^{\alpha\beta}\p_\mu\tilde \chi \p_\nu\tilde \chi
\Bigg)\,.\nn
\ea
In this frame the equations of motion take the following form,
\ba
\label{E_rho}
\mathcal{E}_\rho&=&\tilde{\mathcal{E}}_\pi-\p_\mu \(\det(1-\Sigma)^{\mu\nu}\p_\nu\chi \tilde{\mathcal{E}}_\chi\)=0\\
\mathcal{E}_{\tilde \chi}&=& \det(1-\Sigma)\tilde{\mathcal{E}}_\chi=0\,,
\label{E_tilde chi}
\ea
where $\mathcal{E}_{\tilde \chi}$ is the equation of motion with respect to $\tilde \chi$ in the dual frame, while $\tilde{\mathcal{E}}_\chi$ is the equation of motion \eqref{E_chi} with respect to $\chi$ expressed in terms of the dual variables,
\ba
\mathcal{E}_{\tilde \chi}=\frac{\delta}{\delta \D_s[\chi]} \D_s[\L]\,,\hspace{20pt}
{\rm while}\hspace{10pt}\tilde{\mathcal{E}}_{\chi}=\D_s \left[\frac{\delta}{\delta \chi}  \L\right]\,,
\ea
and similarly for $\mathcal{E}_\rho$ versus $\tilde{\mathcal{E}}_\pi$.
The two sets of equations of motion are not identical but are equivalent.
The equations of motion \eqref{E_chi} and \eqref{E_pi} expressed in terms of the dual variables are
\ba
\label{tilde E chi}
\tilde{\mathcal{E}}_\chi&=&[\(1-\Sigma\)^{-1}]^\mu_{\ \alpha}\p_\mu \left[
[\(1-\Sigma\)^{-1}]^\nu_{\ \beta}\eta^{\alpha \beta}\p_\nu \tilde \chi
\right]+2 g \tilde \chi\(\rho-\frac 1{2 \Lambda^{\sigma}} (\p \rho)^2\)=0\\
\tilde{\mathcal{E}}_\pi&=& \Box \rho+ g  \det(1-\Sigma) \tilde \chi^2=0\,.
\label{tilde E pi}
\ea
Unsurprisingly, the equations of motion \eqref{E_rho} and \eqref{E_tilde chi} are satisfied iff the dual ones \eqref{tilde E chi} and \eqref{tilde E pi} are satisfied. The rest of the argument is thus ran with these two equations $\tilde{\mathcal{E}}_\chi=0$ and $\tilde{\mathcal{E}}_\pi=0$.

\subsubsection*{$\bullet$ Higher derivatives}

The equation $\tilde{\mathcal{E}}_\pi=0$  is a Galileon equation of motion for $\rho$ in which the coefficients depend on $\tilde \chi(x)$. It is manifestly local and contains no more that two time derivatives. The $\chi$ equation of motion $\tilde{\mathcal{E}}_\chi=0$ in \eqref{tilde E chi} is the one that appears to be problematic.
It is manifestly local, but it clearly also includes cubic time derivatives of $\rho$ through the terms symbolically of the form $\Sigma^n \partial  \Sigma(x)\partial  \tilde \chi(x)$. Thus we may be led to believe that we need extra initial data to solve the dynamics. Fortunately this is not the case. Since we have the well defined equation $\Box \rho(x) + g {\rm det}[1-\Sigma(x)] \tilde \chi^2(x) =0 $ this also implies
\be
\frac{\partial}{\partial t} \left(\Box \rho(x) + g {\rm det}[1-\Sigma(x)] \tilde \chi^2(x)  \right)=0 \,.
\ee
This equation may be solved to infer the cubic time derivative of $\rho$ in terms of lower order first and zeroth order time derivatives of $\rho$ and $\tilde \chi$ remembering that we already know from $\Box \rho(x) + g {\rm det}[1-\Sigma(x)] \tilde \chi^2(x) =0\,$ the second order time derivative of $\rho$ in terms of the lower ones. This information may then be substituted back into equation \eqref{tilde E chi} resulting in an equation that determines the second order time derivative of $\tilde \chi$ in terms of lower time derivatives of $\rho$ and $\tilde \chi$. In other words, it still remains true that the initial data needed to solve the dynamics is $\tilde \chi, \rho, \dot{\tilde \chi}$ and $\dot \rho$ and no additional information is needed.

\subsubsection{General argument}

This argument extends to the general matter Lagrangian (\ref{eq2}) however it requires a great deal more work to see it directly from the equations of motion derived from this action.
As in the previous example, it is more straightforward to see it from dualizing the equations of motion. In the general case it works in the same way. The equation of motion for $\pi$ in the original frame never contains more that two time derivatives acting on $\pi$ and one time derivative acting on $\chi$. Since $\Pi$, $\partial \pi$ and $\pi$ all have well defined transformations in terms of $\rho, \partial \rho$ and $\Sigma$ without additional derivatives, and since the transformed form for $\partial_{\mu} \chi$ is $ [\(1-\Sigma\)^{-1}]_{\ \mu}^{\nu}\partial_{\nu} \tilde \chi$ we see than the transformed form of the equation of motion for $\pi$ never contains more than two time derivatives of $\rho$. It may thus be solved for $ \ddot \rho$. Then from (\ref{eq2}) we see that the equation of motion for $\tilde \chi$ never contains more than two derivatives of $\tilde \chi$ but it does contain third derivatives of $\rho$. However these can be inferred as before by differentiating the equation for $\rho$. Together these imply that the combined system can be expressed in a way in which all the equations of motion are second order for the specified general Lagrangian (\ref{eq3}).\\

In fact following these arguments, we could have been more general about our initial choice of Lagrangian for the matter field. We can extend the duality to matter Lagrangians of the form
\ba
S_{\rm matter} = \int \d^d x \, {\cal L}_{\rm matter} ( \chi(x), \partial_{\mu} \chi(x) , \pi(x), \partial_{\mu} \pi(x), \Pi_{\mu \nu}(x)) \, ,
\ea
which is now dual to
\ba
\label{eq2}
\tilde S_{{\rm matter}, s} = \int \d^d x \,  \det\(1-s \Sigma\)  \tilde{\cal L}_{{\rm matter}, s}  \,  ,
\ea
where $\tilde{\cal L}_{{\rm matter}, s}$ is now
\ba
\tilde{\cal L}_{{\rm matter}, s}={\cal L}_{{\rm matter}}\( \tilde \chi, [(1-s \Sigma)^{-1} ]_{\ \mu}^{\nu}\partial_{\nu} \tilde \chi , \rho- \frac{s}{2\Lambda^{\sigma} }(\partial \rho)^2, \partial_{\mu} \rho ,    \left[ \frac{\Sigma}{1-s \Sigma} \right]_{\mu \nu}      \)\,.
\ea
All of these arguments may then easily be generalized to any number of matter fields with arbitrary spin. As long as the matter is treated dynamically, the resulting dual theory is always local if the original theory is local. Similarly the resulting theory will always have second order equations of motion (after the above massaging) if the original theory has the same.

 \subsection{Alternative coupling to matter as an external source}

If one tries to model the matter coupling as an `external' source in the sense of as performed in \cite{deRham:2013hsa} and more recently in \cite{Creminelli:2014zxa}
\ba
S_{\rm int} = \int \d^d x J(x) \pi(x)\,,
 \ea
 then this would map into a non-local coupling \cite{deRham:2013hsa}
 \ba
 \tilde S_{ {\rm int}, s} = \int \d^d x \, {\rm det}(1-s\Sigma)  J(x- s \partial \rho) (\rho - \frac{s}{2\Lambda^{\sigma} }(\partial \rho)^2)\,.
 \ea
 This is however purely a problem with using an external source. A similar problem arises in GR where it is not possible to add an external source for gravity locally without breaking diffeomorphism invariance\footnote{For Schwinger's solution to this in the context of `source theory' see \cite{Schwinger:1968rh}.} but {\it dynamical sources always preserve diffeomorphism invariance}.

   This is pertinent since the duality map is effectively a diffeomorphism, and the failure of the external source to preserve locality is due to the fact that an external source does not transform under diffeomorphisms. However for a dynamical source we can always transform the matter according to \eqref{mattertransform}, \ie  such that there is a dual source $\tilde J(\tilde x) = J(x)$ to make the interaction local\footnote{Here we differ in perspective from \cite{Creminelli:2014zxa}. There it is argued that because the source couples non-locally after the duality transformation the two representations of the coset are inequivalent in their notion of locality. From our point of view this arises because of not accounting for how the matter transforms under the coset. Unlike in internal symmetry coset constructions, when the nonlinearly realized symmetry is an extension of the Poincar\'e group, all matter, which is a function of spacetime, transforms under the additional symmetries of the coset. This is due to the fact that the additional symmetries do not commute with the translation generators of the Poincar\'e group which are used to define the spacetime dependence of matter fields $\chi(x) = e^{-i P.x} \chi(0) e^{i P.x}$. This means that in choosing a different representation of the coset, one must simultaneously choose a different representation for $\pi$ and the matter fields themselves. Equation (\ref{mattertransform}) corresponds to precisely this choice of new representation for the matter fields with the result that the two representations have an equivalent notion of locality.}
   \ba
   \tilde S_{ {\rm int}, s} = \int \d^d x \, {\rm det}(1-s\Sigma)  \tilde J(x) (\rho - \frac{s}{2\Lambda^{\sigma} }(\partial \rho)^2)\,.
   \ea

\subsection{Dual of a Renormalizable theory}

Let us end this section with an example of a (perturbatively) renormalizable theory. Consider the following Lagrangian in four dimensions
\be
S = \int \d^4 x \left[ -\frac{1}{2} (\partial \pi)^2 - \frac{1}{2} (\partial \chi)^2 - \frac{1}{2} m_{\pi}^2 \pi^2- \frac{1}{2} m_{\chi}^2 \chi^2 - \frac{1}{2}g \pi^2\chi^2 - \frac{1}{4!}\lambda_{\pi} \pi^4 - \frac{1}{4!}\lambda_{\chi} \chi^4 \right]   \, .
\ee
This is the most general perturbatively renormalizable Lagrangian for two scalars in 4 dimensions that preserves the discrete symmetries $\pi \rightarrow - \pi$ and $\chi \rightarrow - \chi$.
Following the duality recipe this is dual to
\ba
\label{S renorm}
S_s =&& \int \d^4 x \, {\rm det}(1-s \Sigma)  \left[ -\frac{1}{2}(\partial \rho)^2 - \frac{1}{2}Z_s^{\mu\nu} \partial_{\mu} \tilde \chi \partial_{\nu} \tilde \chi- \frac{1}{2} m_{\pi}^2 \( \rho -  \frac{s}{2\Lambda^3} (\partial \rho)^2\)^2  \right. \\
&&  \left. - \frac{1}{2} m_{\chi}^2 \tilde \chi^2
-\frac{1}{2} g \( \rho -  \frac{s}{2\Lambda^3} (\partial \rho)^2\)^2 \tilde \chi^2 -\frac{1}{4!}\lambda_{\pi}  \( \rho -  \frac{s}{2\Lambda^3} (\partial \rho)^2\)^4   - \frac{1}{4!}\lambda_{\chi} \tilde \chi^4   \right] \,  \nn ,
\ea
where $Z_s^{\mu\nu} = [(1-s \Sigma)^{-1}]^{\mu}_{\ \alpha} [(1-s \Sigma)^{-1}]^{\nu}_{\ \beta} \eta^{\alpha \beta}$. This theory \eqref{S renorm} contains many irrelevant operators, yet it is perturbatively renormalizable if one follows a regularization procedure which is invariant under field redefinitions. For instance dimensional regularization which keeps track only of the logarithmic divergences is invariant under field redefinitions. Following power law divergences might lead to additional operators but power laws depend on the measure of the path integral. More importantly it is well known that taking power law divergences too seriously might lead to erroneous results \cite{Burgess:1992gx} precisely because power law divergences are not invariant under field redefinitions\footnote{It is a common misconception that power law divergences should be kept to see hierarchy problems. On the contrary, all hierarchy problems may be viewed as arising from logarithmic divergences from heavy mass threshold corrections \cite{Burgess:1992gx}. For instance the Higgs hierarchy problem only arises from logarithmic divergences from the assumed beyond the standard model (BSM) physics expected at least at the Planck scale. In the absence of such BSM physics, there is no hierarchy problem.}. This check can be done for instance in QCD where the higher energy theory is known. In these examples following the power law divergences leads to operators which would never have arisen from the higher energy theory. See Ref.~\cite{Burgess:1992gx} for an inspiring discussion on this point.

In summary, following the log divergences that can be trusted since they are independent of the measure and of field redefinitions, then the  \eqref{S renorm} is renormalizable despite including irrelevant operators.

\section{Coupling to Gravity}

\label{sec:Gravity}

We now turn to the important point of how these generalized Galileon theories can be coupled to gravity in a way which preserves the duality symmetry. Within the context of a pure Galileon theory, the coupling of Galileons to gravity was considered in \cite{Deffayet:2009wt} (see also \cite{deRham:2010eu} for its higher dimensional embedding) where the flat spacetime metric $\eta\mn$ is replaced by a curved one $g\mn$ and additional curvature invariants are included to ensure the absence of ghosts. At this level one could be tempted to apply the duality directly at the level of a covariant Galileon and transforming the metric $g\mn$ as a tensor under the coordinate transformation.\\

Whilst tempting, this procedure leads to several conceptual concerns:
\begin{itemize}
\setlength{\itemsep}{0pt}
\setlength{\parskip}{0pt}
\setlength{\parsep}{0pt}
\item First a covariant Galileon does not dualize to another covariant Galileon.

\item Second, as we have emphasized throughout this manuscript, $\pi$ does not transform as a scalar under the duality map. It is rather the derivative of this field that plays that role. As a consequence the Covariant Galileon Lagrangian is not a scalar.

\item  Finally and perhaps most importantly, the Galileon and the duality transformation originated from a genuine gravitational theory in the first place (DGP, massive gravity, bi--gravity). It is therefore only natural to go back to these roots to include the coupling to gravity.
\end{itemize}

Before addressing how to consistently couple the duality to gravity in a way that preserves the duality let us emphasize why the covariant Galileon does not map into itself under the map. Consider a simple example of a covariant Galileon, the cubic Galileon
\ba
\label{S cubic}
S_{\rm cubic} = \int \d^d x  \sqrt{-g(x)}\,\Bigg[  &-& \frac{1}{2} g^{\mu\nu}(x) \partial_{\mu} \pi(x) \partial_{\nu} \pi(x)   \\
&-& \frac{1}{2 \Lambda^{\sigma}}(\Box_{g(x)} \pi(x)) g^{\mu\nu}(x) \partial_{\mu} \pi(x) \partial_{\nu} \pi(x)\Bigg]\,,\nn
\ea
where $\Box_{g(x)}$ is the d'Alembertian for the metric $g_{\mu\nu}(x)$.
Under the duality transformation we have
\be
\partial_{\mu} \pi(x) \rightarrow \tilde \partial_{\mu} \rho(\tilde x) \, .
\ee
The natural choice of transformation for the metric is a diffeomorphism
\be
g_{\mu\nu}(x) = [(1-s \Sigma(\tilde x))^{-1}]_{\ \mu}^{\alpha} [(1-s \Sigma(\tilde x))^{-1}]_{\ \nu}^{\beta} \tilde g_{\alpha \beta}(\tilde x) \, .
\ee
However precisely because $\pi(x)$ and hence $\rho(x)$ does not transform as a scalar we have $\Lambda^\sigma\Sigma_{\mu\nu}(x) = \partial_{\mu}\partial_{\nu} \rho(x) \neq \nabla_{\mu} \nabla_{\nu} \rho(x)$. This means that even on the first term alone the duality generates terms which are not manifestly covariant.

For the first term in \eqref{S cubic} we can compensate this by transforming  the metric instead as
\be
g_{\mu\nu}(x) = \det\(1-s \Sigma(\tilde x)\)^{-2/(d-2)} \tilde g_{\mu \nu}(\tilde x) \, .
\ee
With this choice we have
\be
 \int \d^d x  \sqrt{-g}\,\left[  - \frac{1}{2} g^{\mu\nu} \partial_{\mu} \pi \partial_{\nu} \pi  \right]  \rightarrow  \int \d^d x  \sqrt{-\tilde g}\,\left[  - \frac{1}{2} \tilde g^{\mu\nu} \partial_{\mu} \rho \partial_{\nu} \rho  \right] \, ,
\ee
which is hence duality invariant. However the problem now moves to the cubic term which does not map into a covariant Galileon due to the fact that $(\Box_{g(x)} \pi(x)) $ does not transform as a scalar under this map. This problem is irreconcilable, since $\pi(x)$ is not a scalar under the duality map, and the covariant Galileon assumes $\pi(x)$ is a scalar, then the two ideas are mutually incompatible. An alternative approach which resolves this problem is given in \cite{Gabadadze:2012tr}.

It is worth noting that although the covariant Galileon does not map into itself under the duality, it does map into a consistent local ghost-free theory which is diffeomorphism invariant. However the diffeomorphism invariance is nontrivially realized.

\subsection{A Duality Invariant Gravitational Theory: Massive Gravity}

\label{MGduality}

The origin of the duality was manifest at the level of the decoupling limit of bi--gravity \cite{Fasiello:2013woa}. However the duality can be seen in the even simpler case of Massive Gravity. To see this consider the \stu form of the massive gravity action which can be represented in arbitrary dimensions\footnote{As in GR we may also add Lovelock combinations. However these will not affect any of the subsequent considerations and so we neglect them here.}
\be
S_{\rm MG} =  \mpl^{d-2}\int \d^d x \sqrt{-g} \(\frac{1}{2}R[g]   +  m^2  \sum_{n=0}^d \alpha_n \, {\cal U}_n[\K] \)\,,
\ee
where the tensor $\K\mupn$ is given in terms of the metric and \stu fields by \cite{deRham:2010kj}
\ba
\K\mupn(x) = \delta^{\mu}_{\nu} - \sqrt{g^{\mu \alpha}(x) \partial_{\alpha} \phi^a(x) \partial_{\nu} \phi^b(x) \eta_{ab}}\, .
\ea
This may be viewed as a gravitational theory of a Galileon by choosing the gauge\footnote{Which corresponds to only $d-1$ out of the $d$ allowed gauge choices, the remaining one can be used to fix a gauge for the metric.}
\be
\phi^a(x) = x^a + \frac{1}{\Lambda^{\sigma}} \eta^{ab} \partial_b \pi(x) \, .
\ee
This is a gauge choice one may always choose. It is a useful one since it
implies  that the tensor $\K\mupn$ may be expressed in the form
\be
\K\mupn(x) = \delta^{\mu}_{\nu} - \sqrt{g^{\mu \alpha}(x) (\delta_{\alpha}^a+ \Pi_\alpha^a(x)) (\delta_{\nu}^b+ \Pi_\nu^b(x)) \eta_{ab}} \, ,
\ee
or in matrix notation we have
\be
\K = 1 - \sqrt{g^{-1}(1+\Pi) \eta (1+\Pi^T) } \, .
\ee
Now of course massive gravity includes with it an additional 3 innocuous helicity-one degrees of freedom, but this addition seems to be necessary to maintain the duality.
We now define the duality transformation in the case of a gravitational theory as
\ba
\D_s: \left\{\begin{array}{rcl}
x^\mu & \longrightarrow & \tilde x^{\mu} = x^{\mu} + \frac{s}{\Lambda^{\sigma}}\partial^{\mu} \pi(x) \,  ,\\[5pt]
\p_\mu \pi(x)& \longrightarrow & \tilde \p_\mu \tilde \pi(\tilde x) \, , \\
g_{\mu\nu}(x) & \longrightarrow &  \tilde g_{\mu\nu}(\tilde x)=  [1-s \Sigma(\tilde x)]_{\ \mu}^{\alpha} [1-s \Sigma(\tilde x)]_{\ \nu}^{\beta} g_{\alpha \beta}(x)
\end{array}\right.\,,
\ea
which is to say that the metric transforms in the usual way under the duality diffeomorphism. In matrix form this is the statement that
\be
\tilde g(\tilde x) = [1-s \Sigma(\tilde x)] g(x) [1-s \Sigma^T(\tilde x)]\, .
\ee
We therefore have
\be
\K(x) = 1 - \sqrt{[1-s \Sigma^T(\tilde x)] \tilde g^{-1} [1-s \Sigma(\tilde x)]   \frac{(1+(1-s) \Sigma(\tilde x))}{1-s \Sigma(\tilde x)}\eta  \frac{(1+(1-s) \Sigma^T(\tilde x))}{1-s \Sigma^T(\tilde x)}} \, .
\ee
Now since $\Sigma$ commutes with $\eta$ $(\Sigma_{\mu\nu}= \Sigma_{\nu \mu})$ and using the property of a similarity transformation $S$ that $S \sqrt{A} S^{-1} = \sqrt{S A S^{-1}}$ this is equivalent to
\be
\K(x) = [1-s \Sigma^T(\tilde x)]  \tilde{\K}_s(\tilde x) [1-s \Sigma^T(\tilde x)]^{-1} \,
\ee
where
\be
\tilde{\K}_s = 1 - \sqrt{\tilde g^{-1}(1+(1-s)\Sigma) \eta (1+(1-s)\Sigma^T) } \, .
\ee
However one of the features of the characteristic polynomials is that they are invariant under similarity transformations and so
\be
\, {\cal U}_n[\K(x)] =   \, {\cal U}_n[\tilde{\K}_s(\tilde x)]\,, \hspace{10pt}\forall \hspace{5pt}s\,.
\ee
Furthermore since the metric transforms as a tensor under the duality diffeomorphism we have
\ba
\int \d^d x \sqrt{-g(x)} = \int \d^d \tilde x \sqrt{-\tilde g(\tilde x)}  \, ,
\ea
and
\be
\int \d^d x \sqrt{-g(x)}R[g(x)] = \int \d^d \tilde x \sqrt{-\tilde g(\tilde x)}  R[\tilde g(\tilde x)] \, .
\ee
Substituting into the action and replacing the dummy integration variable $\tilde x$ by $x$ as usual we find that the dual form of the massive gravity action to be
\be
\D_s:  S_{\rm MG} \longrightarrow \tilde{S}_{{\rm MG}, s} = \mpl^{d-2} \int \d^d x \sqrt{-\tilde g} \( \frac{1}{2} R[\tilde g]   +  m^2  \sum_{n=0}^d \alpha_n \, {\cal U}_n[\tilde{\K}_s] \) \,,
\ee
where
\be
\tilde{\K}_s{}\mupn(x) = \delta^{\mu}_{\nu} - \sqrt{\tilde{g}^{\mu \alpha}(x) (\delta_{\alpha}^a+ (1-s)\Sigma_\alpha^a(x)) (\delta_{\nu}^b+ (1-s) \Sigma_\nu^b(x)) \eta_{ab}} \, .
\ee
However this is nothing other than the original \stu form of the massive gravity action in which we have chosen the gauge for the \stu fields to be
\be
\phi^a(x) = x^a + (1-s) \frac{1}{\Lambda^{\sigma}} \eta^{ab} \partial_b \rho(x) \, .
\ee
Thus the duality transformation simply transforms between a one-parameter family of gauge choices. In particular, one of these choices $s=1$ is none other than unitary gauge
\be
\phi^a_{s=1, \rm unitary }(x) = x^a \, .
\ee
Thus we see that the \stu form of the massive gravity Lagrangian is manifestly invariant under the duality because the duality is nothing other than a change of diffeomorphism gauge.  This also allows us to understand why the matter had to transform in the chosen way. Any matter that couples locally to the metric $g_{\mu \nu}(x)$ will remain invariant under the transformation provided that matter fields transform under the associated diffeomorphism as we advocated in section~\ref{mattercoupling}.\\

These arguments easily extend to bi--gravity, and in fact are already implicit in the derivation of the duality from bi--gravity given in \cite{Fasiello:2013woa}. The only price we have paid in coupling the Galileon to gravity in this way is that we need to also introduce an additional helicity-one degrees of freedom. However, we may note that the dynamics of the helicity-one, which can be captured in the decoupling limit action \cite{Ondo:2013wka} (see also  \cite{Gabadadze:2013ria}), are rather tame. Indeed it is always consistent to set the helicity-one mode to zero classically since if matter couples to the metric covariantly it does not get sourced.\\

Recognizing that the duality transformation is little more than a change of gauge is also crucial to understanding that although the duality map appears to be non-local, it completely preserves the notion of locality since the definition of locality does not depend on the gauge choice. Furthermore this also gives an independent argument of why the duality should remain true at the quantum level since it could only be violated in a theory with a diffeomorphism anomaly \cite{AlvarezGaume:1983ig} which has yet to occur in four dimensions.

\subsection{A New Class of Duality Invariant Massive Gravity Theories}

We have seen above that the massive gravity action of \cite{deRham:2010kj} is manifestly invariant under the duality symmetry. It is also by now well known that the decoupling limit of massive gravity corresponds to a Galileon theory \cite{deRham:2010ik}. This begs the question, {\it Is there a generalization of the massive gravity action for which the decoupling limit corresponds to the generalized Galileons~?} One approach to this question is given in \cite{Gabadadze:2012tr} where gravity is coupled to an extra scalar field in a way that preserves the nonlinearly realized symmetries acting on that scalar. Here we would like to take a different approach where the `Galileon' field remains as the helicity-zero scalar of a massive graviton.\\

As we have emphasized above, the nonlinearly realized duality is built into massive gravity theories and gets linearly realized when we work with the diffeomorphism invariant \stu formulation. With this in mind we may ask whether there are other interactions we can allow with the \stu fields that preserve the following criteria:
\begin{itemize}
\setlength{\itemsep}{0pt}
\setlength{\parskip}{0pt}
\setlength{\parsep}{0pt}
\item Global Lorentz invariance $\phi^a \rightarrow \Lambda^a{}_b \phi^b$ , with $\Lambda \eta \Lambda^T = \eta$ ,
\item Locality,
\item Diffeomorphism invariance,
\item Absence of Boulware-Deser ghost \cite{Boulware:1973my}.
\end{itemize}

Locality requires that the Lagrangian is only a function of $\phi^a$ and first derivatives of $\phi^a$. The absence of the Boulware-Deser ghost requires that the dependence of the Lagrangian on derivatives of the \stu fields occurs through the characteristic polynomials given in \cite{deRham:2010kj}.

\subsubsection{New \stu dependence}

These arguments lead to the following generalization of the massive gravity action
\ba
S_{\rm MG} &=&\mpl^{d-2} \int \d^d x \sqrt{-g}\Bigg[ \frac{1}{2} \Phi(\phi^a \phi_a)R[g]   +  m^2  \sum_{n=0}^d \alpha_n(\phi^a \phi_a) \, {\cal U}_n[\K]   \nn \\
&+& \sum_{n=2}^{[d/2]} \,  \Phi_n(\phi^a \phi_a) {\cal L}^{(n)}_{\rm Lovelock}[g]\Bigg] \, .
\ea
Here ${\cal L}_{\rm Lovelock}$  are the usual Lovelock Lagrangians (Gauss-Bonnet in four dimensions). We have now promoted the Planck mass ($\mpl^{d/2}\Phi$), mass parameters ($\alpha_n$), and the Lovelock Lagrangian
coefficients ($\Phi_n$) to be functions of the Lorentz invariant combinations
of the \stu fields\footnote{If one were interested in massive gravity on another reference metric, for instance (Anti)-de Sitter we could easily extend the relation \eqref{phi square} appropriately.}
\ba
\label{phi square}
\phi^a \phi_a = \phi^a(x) \phi^b(x) \eta_{ab} \, .
\ea
To see that this is a gravitational theory with the Galileon duality we can choose the gauge
\be
\phi^a(x) = x^a + \eta^{ab} \partial_b \pi(x) \, .
\ee
Following the previous arguments this action respects the duality symmetry in the form
\ba
\D_s: \left\{\begin{array}{rcl}
x^\mu & \longrightarrow & \tilde x^{\mu} = x^{\mu} + \frac{s}{\Lambda^{\sigma}}\partial^{\mu} \pi(x) \,  ,\\[5pt]
\p_\mu \pi(x)& \longrightarrow & \tilde \p_\mu \tilde \pi(\tilde x) \, , \\
g_{\mu\nu}(x) & \longrightarrow &  \tilde g_{\mu\nu}(\tilde x)=  [1-s \Sigma(\tilde x)]_{\ \mu}^{\alpha} [1-s \Sigma(\tilde x)]_{\ \nu}^{\beta} g_{\alpha \beta}(x) \, .
\end{array}\right.
\ea
We may also choose to define $\uppi$ via
\be
\phi^a(x) =  \eta^{ab} \partial_b \uppi(x) \, ,
\ee
so that the duality transformation becomes
\ba
\D_s: \left\{\begin{array}{rcl}
x^\mu & \longrightarrow & \tilde x^{\mu} = \frac{s}{\Lambda^{\sigma}}\partial^{\mu} \uppi(x) \,  ,\\[5pt]
&& x^{\mu} = - \frac{s}{\Lambda^\sigma} \tilde \partial^{\mu} \uprho(\tilde x) , \\
g_{\mu\nu}(x) & \longrightarrow &  \tilde g_{\mu\nu}(\tilde x)=  [s \Upsigma(\tilde x)]_{\ \mu}^{\alpha} [s \Upsigma(\tilde x)]_{\ \nu}^{\beta} g_{\alpha \beta}(x) \, .
\end{array}\right.
\ea
The two are related by the shift $\uppi =\pi + \frac{\Lambda^{\sigma}}{2s }x^{\mu}x_{\mu}$.  \\

Using this second form the action can be expressed as
\ba
S_{\rm MG} &=&\mpl^{d-2} \int \d^d x \sqrt{-g} \Bigg[ \frac{1}{2} \Phi\(\frac{1}{\Lambda^{2\sigma}}(\partial^{a} \uppi(x))^2\)R[g]   \nn \\
 &+& m^2  \sum_{n=0}^d \alpha_n\(\frac{1}{\Lambda^{2\sigma}}(\partial^{a} \uppi(x))^2\) \, {\cal U}_n\left[1- \sqrt{g^{-1}\Uppi \eta  \Uppi^T}\right]    \nn \\
&+& \sum_{n}   \Phi_n\(\frac{1}{\Lambda^{2\sigma}}(\partial^{a} \uppi(x))^2\) {\cal L}^{(n)}_{\rm Lovelock}[g] \Bigg]\, ,
\ea
which is clearly a gravitational extension of the generalized Galileon actions in which there is assumed to be no explicit dependence on $\uppi$, only on first and second derivatives of $\uppi$. That is to say that this action is a gravitational extension of the generalized Galileon which preserves the shift symmetry $\uppi \rightarrow \uppi + c$.

Now this is a new class of massive gravity theories that preserves Lorentz invariance at the price of breaking translation invariance for the vacuum. This is explicit in unitary gauge $\phi^a=x^a$ where we have
\ba
\label{unitary}
S_{\rm unitary} &=&\mpl^{d-2} \int \d^d x \sqrt{-g}\Bigg[ \frac{1}{2} \Phi\(x^2\)R[g]   +
  m^2 \sum_{n=0}^d \alpha_n\(x^2\) \, {\cal U}_n\left[1- \sqrt{g^{-1} \eta}\right]   \nn \\
&+& \sum_{n}       \, \Phi_n\(x^2\) {\cal L}^{(n)}_{\rm Lovelock}[g] \Bigg]\, ,
\ea
with $x^2 = x^a x^b \eta_{ab}$.
This explicit dependence  on $x^ax_a$ implies that translation invariance is broken but not Lorentz invariance, at least around the preferred point $ x^a=0$. For this reason the linearized theory around the vacuum does not take the usual Fierz-Pauli form which explains why this form of the Lagragian was not recognized before.

\subsubsection{Degrees of freedom count}

$\bullet$ {\bf Primary Constraint}

Let us now perform the count on the number of degrees of freedom of these generalized massive gravity theories. To do this, rather than using the unitary gauge Lagrangian (\ref{unitary}) we shall utilize a non-Lorentz invariant  yet perfectly acceptable version of unitary gauge for which
\be
\phi^0 = \sqrt{t^2 + \vec x^2}  \quad {\rm and}\quad  \phi^i = x^i\,,
\ee
so that $\phi^a \phi_a =- t^2$, so that in this case the reference metric is Minkowski in a non-standard coordinate system
\be
f_{\mu\nu} \d x^{\mu} \d x^{\nu} = \partial_{\mu} \phi^a \partial_{\nu} \phi^b \eta_{ab} \d x^{\mu} \d x^{\nu}  \, .
\ee
This gauge choice may be made in any region for which $\phi^a \phi_a <0$ since we have $\phi^a \phi_a = -t^2$. This breaks manifest Lorentz invariance but has the virtue of removing any complication with respect to integrating by parts spatial derivative terms which enter into the calculation of the Poisson brackets needed to probe the constraints.
Since the number of degrees of freedom is independent of the gauge used to perform the analysis then the following analysis should also go through in the original Lorentz invariant unitary gauge. In making this argument we are implicitly analytically continuing from the region $\phi^a \phi_a >0$ to the region $ \phi^a \phi_a \le 0$. Again we would argue that the number of degrees of freedom cannot change on this analytic continuation. We leave to future work a detailed derivation of the degrees of freedom in the Lorentz invariant unitary gauge.

As usual it is helpful to go to the ADM phase space form. On doing so the Lagrangian takes the schematic form (for simplicity we focus on the case without Lovelock terms)
\be
S_{\rm unitary} = \int  \d^d x \left[  \pi^{ij} \dot g_{ij} - {\cal H}(N,N^i, g_{ij}, \partial_i g_{jk}, \pi_{ij},t^2, \vec x)  \right]\,.
\ee
where $\pi^{ij}$ is the momentum conjugate to 6 components of the spatial part of the metric $g_{ij}$.
The new feature relative to the usual massive gravity case is that the Hamiltonian density has an explicit dependence on $t$ through the Planck mass and other mass parameters.
Nevertheless  there still exists a primary constraint due to the vanishing of the Hessian
\be
{\rm det} \left[ \frac{\partial^2  {\cal H}}{\partial N^{\mu} \partial N^{\nu}}\right]=0 \,,
\ee
with $N^{\mu} = \{N, N^i\}$. The presence and the form of this constraint is unchanged by the presence of the $t^2$ dependence because the above equation contains no partial derivatives with respect to time. It is this constraint that removes (one half of) the Boulware-Deser ghost.

This constraint may then be made manifest by solving
\be
\frac{\partial}{\partial N^i }{\cal H}(N,N^i, g_{ij}, \partial_i g_{jk}, \pi_{ij},t^2, \vec x) =0 \, ,
\ee
for $N^i$ and substituting back in. That this is possible can be inferred by continuity with the usual massive gravity where it has been shown in \cite{Hassan:2011hr}.

Once this is done the action takes the form
\be
S_{\rm unitary} = \int  \d^d x \left[  \pi^{ij} \dot g_{ij} - {\cal H}_0(g_{ij}, \partial_i g_{jk}, \pi_{ij},t^2, \vec x) - N C_1(g_{ij}, \partial_i g_{jk}, \pi_{ij},t^2, \vec x) \right]\,,
\ee
where the Boulware-Deser ghost removing constraint $C_1=0$ is now enforced by the Lagrange multiplier $N$.\vspace{10pt}

\noindent $\bullet$ {\bf Secondary Constraint}

Defining the Hamiltonian by
\be
\mathfrak{H}_1 = \int \d^{d-1} x \,  {\cal H}_0(g_{ij}, \partial_i g_{jk}, \pi_{ij},t^2, \vec x)\,,
\ee
and the integrated constraint by
\be
\mathfrak{C}_1 = \int \d^{d-1} x  \, N C_1(g_{ij}, \partial_i g_{jk}, \pi_{ij},t^2, \vec x)\,,
\ee
then to show that there is a secondary constraint we would have to compute the usual Poisson brackets
\be
\frac{\partial}{\partial t} C_1 = \frac{\partial^{\rm exp} C_1}{\partial {t}} + \{ C_1, \mathfrak{H}_1 \} + \{ C_1, \mathfrak{C}_1 \} \sim 0  \, ,
\ee
where $\partial^{\rm exp}/{\partial t}$ refers to differentiation with respect to the explicit time dependence in $C_1$ through the Planck mass and mass parameters. \\

The only new feature relative to the usual massive gravity proof is that $ \frac{\partial^{\rm exp} C}{\partial t}  \neq 0$. However the existence of a secondary constraint requires only that this equation cannot be viewed as an equation for the lapse $N$. But the lapse only enters in the term $\{ C_1, \mathfrak{C}_1 \} $ which up to an $t^2$ dependence of the mass parameters is identical to what it is in the usual massive gravity case (for a non Cartesian reference metric). Thus we may immediately borrow the proof that $\{ C_1, \mathfrak{C}_1 \} \sim 0 $  from Ref.~\cite{Hassan:2011hr} where this is shown for any reference metric. This follows because the Poisson bracket computation is immune to the time-dependence and the only difference is that the Planck mass and mass parameters are dependent of time.  \\

Given that the secondary constraint exists with $C_2=\frac{\partial^{\rm exp} C_1}{\partial t} + \{C_1, \mathfrak{H}_1 \}$, this may in turn be included back into the Lagrangian utilizing a second Lagrange multiplier $\mu$
\ba
S_{\rm unitary} &=& \int  \d^d x \left[  \pi^{ij} \dot g_{ij} - {\cal H}_0(g_{ij}, \partial_i g_{jk}, \pi_{ij},t^2, \vec x) \right.  \nn \\
&& \left. - N C_1(g_{ij}, \partial_i g_{jk}, \pi_{ij},t^2, \vec x) - \mu C_2(g_{ij}, \partial_i g_{jk}, \pi_{ij},t^2 , \vec x)\right]\,.
\ea
%\vspace{10pt}

\noindent $\bullet$ {\bf Tertiary Constraint}

Next we must check for the existence of a tertiary constraint. For this we must compute
\be
\frac{\partial}{\partial t} C_2 = \frac{\partial^{\rm exp} C_2}{\partial {t}} + \{ C_2, \mathfrak{H}_1 \} + \{ C_2, \mathfrak{C}_1 \}+ \{ C_2, \mathfrak{C}_2 \} \,,
\ee
where
\be
\mathfrak{C}_2 = \int \d^{d-1} x  \, \mu C_2(g_{ij}, \partial_i g_{jk}, \pi_{ij},t^2, \vec x)  \, .
\ee
We would only generate a tertiary constraint if it were not possible to view this as an equation that determines one of the Lagrange multipliers $N$ or $\mu$. However we already know that this is possible in the normal massive gravity case, and so by continuity this equation must always be solvable for the Lagrange multipliers. Thus there can be no tertiary constraint.

In summary we necessarily obtain two second class constraints $C_1, C_2$ which are sufficient to remove the Boulware-Deser ghost. Consequently these generalized massive gravity theories that preserve Lorentz invariance at the price of spontaneously breaking translation invariance ($P |0 \rangle \neq 0$) contain 5 propagating degrees of freedom. Finally the fact that these theories are continuous in theory space with the usual massive gravity Lagrangian, and contain no new degrees of freedom,   justifies them being viewed as generalizations of massive gravity.

\section{Causality and (super)luminality}
\label{sec:SL}

We now turn to the important question of causality in this class of theories.
In a relativistic quantum theory, relativistic causality requires that all local fields commute at space-like separations.
\be
\label{causality}
[{\cal O}(x), {\cal O}(y) ] =0 \, , \, \text{for }\, (x-y)^2>0 \, .
\ee
A necessary condition for this to occur is that the {\bf front velocity} (sometimes called wavefront) is luminal. However the condition (\ref{causality}) can be satisfied when both the group and phase velocities at low energies are superluminal.  Thus neither the group nor phase velocity (at low energies) being superluminal are indicative of acausality.

Since this distinction will be crucial to our subsequent analysis, we sketch the essential details of this proof of why it is the front velocity that determines the causal structure of a quantum theory.

\subsection{Front velocity and causality}

Consider a typical quantum system whose fluctuations in a non-vacuum state $| \alpha \rangle $ satisfy the dispersion relation $\omega=\omega(k)$ where we assume rotational invariance for simplicity $k= | \vec k|$. The retarded propagator for the system in a state $|\alpha \rangle$ can be defined by
\be
G_{\rm ret}(x,x') = - i \theta(t-t') \langle \alpha |[{\cal O}(x), {\cal O}(y) ]  | \alpha \rangle\, .
\ee
Although this vanishes for $t<t'$ as required for causality, for this notion of causality to be Lorentz invariant we require also that $[{\cal O}(x), {\cal O}(y) ] =0 $ for spacetime separations so that the support of the retarded propagator lies entirely in the future lightcone.

Focusing on $d=4$ the retarded propagator typically takes the form
\be
G_{\rm ret}(x,0) = -i \theta(t)  \int \frac{\d^3 k}{(2 \pi)^3 2 \omega(k)} \left(  e^{i \vec k. \vec x - i \omega(k) t} -  e^{i \vec k. \vec x + i \omega(k) t} \right) \, .
\ee
Performing the angular $k$ integrals we have
\be
G_{\rm ret}(x,0) = -\theta(t)  \int_0^{\infty} \frac{\d k}{(2 \pi)^2 2 \omega(k)} \frac{k}{r} \left(  e^{ - i \omega(k) t} -  e^{ + i \omega(k) t} \right)(e^{ikr}- e^{-ikr}) \, .
\ee
which may be reexpressed as
\be
G_{\rm ret}(x,0) = -\theta(t)  \int_{-\infty}^{\infty} \frac{\d k}{(2 \pi)^2 2 \omega(k)} \frac{k}{r} \left(  e^{ - i \omega(k) t} -  e^{ + i \omega(k) t} \right) e^{ikr} \, .
\ee
The time-dependent oscillatory factors can be viewed as coming from an integral of the form
\be
G_{\rm ret}(x,0) = i\theta(t)  \int_{-\infty}^{\infty}  \d E \int_{-\infty}^{\infty} \frac{\d k}{(2 \pi)^3 } \frac{k}{r}
\frac{e^{ - i E t}e^{ikr}}{\omega(k)^2 - (E+ i \epsilon)^2}  \, ,
\ee
which makes manifest the famous $i \epsilon$ prescription that determines the retarded propagator. In fact in this form we can do away with the theta function $\theta(t)$, however we keep it for clarity.

Next we rewrite the dispersion relation $\omega = \omega(k)$ in terms of a refractive index $n(\omega)$ via $k = n(\omega) \omega$ and defining the refractive index for negative $\omega$ to be $n(-\omega) = n(\omega)$  for $\omega$ real) then we can rewrite this integral as
\ba
G_{\rm ret}(x,0) = i\theta(t)  \int_{-\infty}^{\infty}  \d E \int_{-\infty}^{\infty} \frac{\d \omega }{(2 \pi)^3 } \frac{n(\omega) \omega}{r} \left( n(\omega) + \omega \frac{\d n(\omega)}{\d \omega}\right) \frac{ e^{ - i E t}e^{i n(\omega) \omega r}}{\omega^2 - (E+ i \epsilon)^2} \,.\ \
\ea
The first key point is to assume that $n(\omega)$ and in particular $e^{i n(\omega) \omega r}$ are analytic functions in the upper-half complex $\omega$ plane\footnote{To be clear, by analytic we also mean at infinity itself, so we also assume that ${\rm Im}[n(\omega)\omega]>0$ as $|\omega|\to \infty$ in the upper half of the complex $\omega$ plane.}. If this is true then we can close the contribution of the $\omega$ integral with an infinite semi-circle in the upper-half complex plane. By Cauchy's theorem, the only contribution to the integral comes from the only pole that lies in the upper-half plane, namely $\omega = E_+=E+ i \epsilon$.

The residue of this pole gives us
\be
G_{\rm ret}(x,0) =- \theta(t)  \int_{-\infty}^{\infty}  \frac{\d E}{8 \pi^2 } \frac{n(E) }{r} \left( n(E) + E\frac{\d n(E)}{\d E}\right)  e^{ - i E \(t-n(E) r\)} \, .
\ee
Physically this is the statement that the influence of a disturbance at the origin will expand out in an outgoing spherical wave with phase velocity $\frac{\omega}{k}=1/n(\omega)$ (had we not assumed spherical symmetry we would have additional outgoing multipoles). \\

Since the Fourier transform of any function which vanishes for $t<0$, \ie $G_{\rm ret}(x,0)$,  must be analytic in the upper-half complex energy $E$ plane, this justifies our assumption that the refractive index $n(E)$ and $e^{i n(E) E r}$ can be extended to analytic functions in the upper-half complex energy plane. This is the famous connection between causality and analyticity that is used to argue for analyticity of the S-matrix. The connection is that a particle travelling through a medium acquires a nontrivial refractive index through scatterings off the particles that compose the medium. Hence the analyticity of the refractive index is directly determined by the analyticity of the S-matrix. In the present case the medium is typically a background vev for the scalar field encoded in the non-vacuum state $| \alpha \rangle$.\\

Now to the second crucial point: Provided that $n(E) \rightarrow 1$ as $|E| \rightarrow \infty$ (at least in the upper half of the complex $E$ plane, including the real axis) , then for $r>t$ the integral may also be evaluated by performing a contour integral along the real axis which is closed by an infinite radius semi-circle in the upper-half complex energy plane. This is true because as $|E| \rightarrow \infty$ the exponential terms are
\be
\lim_{|E| \rightarrow \infty\   {\rm Im} [E] \ge 0} e^{ - i E \(t-n(E) r\)}  \sim e^{ iE (r-t)}\,,
\ee
which are well behaved for ${\rm Im} [E] >0$ assuming $r>t$. Since the refractive index $n(E)$ and $e^{i n(E) E r}$  are analytic functions in the upper-half complex $E$ plane there are no poles or branchcuts to deal with and so Cauchy's theorem tells us that
\be
G_{\rm ret}(x,0) =0  \, \quad {\rm for } \quad r>t  \equiv x^2>0\, .
\ee
This is precisely the condition required for relativistic causality. On the contrary for $r<t$ the integrals would have to be deformed in the lower-half plane where the integrand is not analytic, and a nonzero result would be obtained.\\

The two key assumptions in this argument are:
\begin{enumerate}[(a)]
\setlength{\itemsep}{0pt}
\setlength{\parskip}{0pt}
\setlength{\parsep}{0pt}
\item  Analyticity of the refractive index in the upper-half complex energy plane\footnote{It is an elementary result that any function defined on the real axis that vanishes or approaches a constant at infinity sufficiently rapidly can be extended to an analytic function in the upper-half complex plane. } and
\item  The fact that $n(\omega) \rightarrow 1 $ as $\omega \rightarrow \infty$.
\end{enumerate}
Since the phase velocity is defined as
\be
v_{\rm phase}(k) = \frac{\omega}{k}= \frac{1}{n(\omega)} \,
\ee
the requirement (b) is that the infinite $\omega$ (\ie infinite momenta) limit of the phase velocity, a.k.a. the {\bf front velocity} is luminal
\be
v_{\rm front} = \lim_{k \rightarrow \infty} v_{\rm phase}(k) \rightarrow 1 \, .
\ee
In other words {\bf relativistic causality is synonymous with the front velocity being luminal}.\\

In particular we see that the group velocity $v_{\rm group} = \frac{\d \omega(k)}{\d k}$ does not even enter into the discussion of causality. Thus there is no problem for the group velocity and phase velocity for finite $k$ to be  superluminal. Experiments have now confirmed the reality of superluminal group velocities in optical systems (see Ref.~\cite{Milonni} for discussions on the physical context). Indeed the fact that superluminal group velocities are not in conflict with causality was understood and resolved by Sommerfeld and Brillouin \cite{Brillouin}. \\

Despite these known and old results, there is a commonly stated folk theorem that `a low energy effective field theory with superluminal propagation admits no Lorentz invariant UV completion'. This folk theorem is violated by the real world. In addition to the known optical systems which admit superluminal group velocities at low energies \cite{Milonni}, it can also be shown that the low energy effective field theory of photons in a curved spacetime admits superluminal group and phase velocities and that this is not in conflict with the fact that its UV completion (QED in a curved spacetime) is Lorentz invariant (see \cite{Shore:2003zc} for a review and \cite{Shore:2007um,Hollowood:2007kt}). \\

One of the arguments for the folk-theorem is the Kramers-Kronig dispersion relation which is a consequence of analyticity. It states that, given the refractive index is analytic in the upper half complex plane then from Cauchy's theorem we have
\be
n(0) - n(\infty) = \frac{1}{ i \pi} P \left[ \int_{-\infty}^{\infty} \frac{\d \omega}{\omega}  {n(\omega)} \right]=  \frac{1}{ i \pi} \int_{0}^{\infty} \frac{\d \omega}{\omega}  {\rm Disc}  \left[n(\omega) \right]\,,
\ee
where
\be
 {\rm Disc}  \left[n(\omega) \right] = n(\omega+i \epsilon) - n(- \omega +i \epsilon).
\ee
Then assuming the normal hermitian analyticity condition $n(\omega^*) = n^*(\omega)$, this would imply
\be
n(0) - n(\infty) = \frac{2}{ \pi}   \int_{0}^{\infty} \frac{\d \omega}{\omega} {\rm Im} \left[ n(\omega) \right] \, .
\ee
Since the optical theorem determines the imaginary part of the refractive index ${\rm Im}[n(\omega)]$ to be positive then we must have $n(\infty) \le n(0)$ which implies $v_{\rm phase}(0) \le v_{\rm front}$. This in turn would seem to imply that if superluminalities are found at low energies, then the front velocity must be superluminal hence violating Lorentz invariance in the UV. \\

These arguments are violated in the low energy effective theory of photons in a curved spacetime despite the fact that this theory admits a UV completion (see
\cite{Hollowood:2007kt,Hollowood:2007ku,Hollowood:2008kq,Hollowood:2009qz,Hollowood:2010bd,Hollowood:2010xh,Hollowood:2011yh,
Hollowood:2012as}  for a full discussion). The resolution \cite{Hollowood:2008kq} is that although the refractive index is analytic in the upper half complex energy plane, as required for causality, it does not satisfy the condition of hermitian analyticity. This is a condition that usually arises in S-matrix theory in Minkowski that is not justified on the grounds of causality alone. In a nonzero background there is no expectation that this condition must hold. Furthermore the local value of ${\rm Im} \left[ n(\omega) \right] $ can become negative without violating the optical theorem since the optical theorem holds only globally \cite{Hollowood:2012as}. If we view the Galileon theories in the context of how they arise in decoupling limit of DGP/massive gravity/bi--gravity etc., then a non-zero vev for the Galileon is tantamount to working with a curved spacetime. Thus the usual analyticity assumptions assumed for scattering for non-gravitational theories do not apply. All that is required is the analyticity in the upper half complex energy plane required for relativistic causality. \\

By the very definition of the front velocity, lying at infinite momenta, it simply  does not make sense to perform a pure classical calculation to infer $v_{\rm front}$. The front velocity requires knowledge of the full quantum corrected correlation functions. In Galileon type theories and in massive gravity, we cannot trust a tree level calculation of the front velocity precisely because perturbation theory breaks down at the scale $k \sim \Lambda$ (or the appropriate redressed scale). In particular that means that if we use perturbation theory to compute $v_{\rm phase}(k=\Lambda)$ we find order unity corrections (if the background has order unity Lorentz violation). Thus it is impossible to anticipate $v_{\rm front}$ from a perturbative calculation.  \\

%In what follows we shall show how Galileon can have superluminal group velocity and yet lead to no problem with causality.

\subsection{Superluminalities in Galileons?}

The presence of apparent superluminalities in Galileon and related theories was first pointed out in Refs.~\cite{Adams:2006sv,Nicolis:2008in,Hinterbichler:2009kq}  and later by many other subsequent equivalent analyses. All of these analyses and subsequent work have relied on a classical (tree level) calculation which breaks down precisely when the relation with causality is considered (see Ref.~\cite{Burrage:2011cr} for more details).  See also \cite{deRham:2014zqa} for a review on massive gravity and related superluminality considerations. \\

%
%In all the examples provided in the literature  what has been computed is the classical group velocity which in these examples turns out to be equivalent to the phase velocity at the classical level and the {\it classical front velocity} despite the fact that quantum corrections to this classical front velocity dominate. More precisely, these calculations effectively compute the retarded propagator or the front velocity {\bf at tree level only}.

As we discussed above in models such as the Galileon, perturbation theory breaks down at the low scale $\Lambda$ (or the redressed scale depending on the background). Consequently the tree level computation of the phase and group velocities can only be trusted for $k \ll \Lambda$. Thus while superluminalities have been found in these calculations,  they do {\bf not} imply that the front velocity is superluminal, since that requires an understanding of the UV theory $k \rightarrow \infty$ and therefore they {\bf do not imply any acausalities}.\\

What is needed in Galileon models, is a UV description which is capable of dealing with the regime $k > \Lambda$. A possible such UV description with be described elsewhere \cite{Luke} and we will present arguments that the front velocity is luminal for precisely those solutions where superluminal group velocities have been found.\\

An independent criticism of Galileon theories was given in \cite{Adams:2006sv} which argued that for Galileons the S-matrix could not satisfy its usual analyticity properties. In the context of the above argument this corresponds to saying that the refractive index $n(\omega)$ may not be an analytic function in the upper-half complex $\omega$ plane which was our additional assumption needed to prove relativistic causality. However as we have already discussed it is possible that a theory may not satisfy all the historical analyticity assumptions of the S-matrix, but still satisfies the physically required condition of analyticity of the refractive index in the upper half plane. Furthermore, we will argue elsewhere that these perturbative arguments have not fully accounted for the role of the Vainshtein mechanism \cite{Vainshtein:1972sx} at the quantum level \cite{Luke} and that with the benefit of a non-perturbative UV description of Galileons one may argue that the S-matrix satisfies the necessary analytic properties for causality (see also \cite{Cooper:2013ffa} for other related theories and considerations).\\

For now we shall content ourselves with understanding how the duality affects the causal structure in the classical theory with the caveat that this is at best a description of the low energy physics and thus is unlikely to correspond to the actual causal behaviour of the quantum theory.

\subsection{Example with superluminal classical group velocity}

We now turn to how the duality affects the question of classical (super)luminality. To understand this let us go back to the  simple example of the interacting two field system studied in section \ref{sec:example}
\ba
\label{EG}
S = \int \d^d x  \left[ -{\rm det}(1+ \Pi) \frac{1}{2}(\partial \pi)^2- \frac{1}{2} (\partial \chi)^2 + g \chi^2 \pi \right] \, .
\ea
In the $\pi$-duality frame the equations of motion are given by $\E_\chi=0$ and $\E_\pi=0$ in \eqref{E_chi} and \eqref{E_pi} while in the $\rho$-duality frame they are given by $\tilde{\E}_\chi=0$ and $\tilde{\E}_\pi=0$ in \eqref{tilde E chi} and \eqref{tilde E pi}.
%
%
%for which the equations of motion in the $\pi$-duality frame are
%\ba
%&& \Box \chi + 2 g \pi \chi =0 \, , \\
%&& {\rm det}[1+\Pi] {\rm Tr}\left[\frac{\Pi}{1+\Pi}\right] + g \chi^2 =0 \, .
%\ea
%and in $\rho$-duality frame are
%\ba
%&& \eta^{\mu \alpha }[(1- \Sigma(x))^{-1} ]_{\ \mu}^{\nu}\partial_{\nu} \left( [(1- \Sigma(x))^{-1} ]_{\alpha}^{\beta}\partial_{\beta} \tilde \chi(x) \right) + 2 g \left(\rho(x) + \frac{1}{\Lambda^\sigma} (\partial \rho(x))^2 \right) \tilde \chi(x) =0 \, \nn \\
%&& \Box \rho(x) + g \tilde \chi(x)^2 {\rm det}[1-\Sigma(x)] =0\, .
%\ea
As we have emphasized, both systems of equations are local and second order in time.

\subsubsection*{$\bullet$ Classical velocity in the $\pi$-duality frame}

Now in the $\pi$-duality frame, $\chi$ always propagates at the speed of light even around an arbitrary background. To see this we consider fluctuations about the arbitrary background given by $\bar \pi$ and $\bar \chi$. Expressing the fields as $\pi=\bar \pi+\delta \pi$ and $\chi=\bar \chi+\delta \chi$, the linearized equation of motion for $\delta \chi$ about this background is then
\ba
 \Box \delta \chi + 2 g \bar \chi \delta \pi+ 2 g \bar \pi \delta \chi =0 \,,
\ea
and so in this frame $\chi$ always propagates at the speed of light about any background given by $\bar \pi$ and $\bar \chi$.

On the other hand we can straightforwardly find background solutions for which the fluctuations of $\pi$ are {\bf classically} superluminal. To see this we note that the equation of motion for $\delta \pi$ when expanded around a background for which $\bar \chi=0$ takes the form
\ba
 \partial_{\mu} \left( Z^{\mu \nu} \partial_{\nu} \delta \pi \right)=0\,,
\ea
with
\ba
Z\mupn &=&\det\(1+\bar \Pi\) \(\left[ (1+ \bar \Pi)^{-2} \right]\mupn +{\rm Tr}\left[ \frac{\bar \Pi}{1+\bar \Pi}\right] \left[(1+\bar \Pi)^{-1}\right]\mupn\)\\
&=&
\det\(1-\bar \Sigma\)^{-1}\( \left[ (1- \bar \Sigma)^2 \right]\mupn +{\rm Tr}[\bar \Sigma] [1-\bar \Sigma]\mupn\)\,.
\ea
Without loss of generality we now perform a Lorentz transformation in the vicinity of the point $x$ so that $\bar \Pi$ and hence $\bar \Sigma$ are both diagonal at that point.
% Then the time eigenvalue of $Z$ is given by
%\be
%Z^0_0 \propto (1-\bar \Sigma^0_0) ( 1+ \bar \Sigma^1_1+\bar \Sigma^2_2+\bar \Sigma^3_3)
%\ee
%and the space eigenvalue of $Z$ by
%\be
%Z^1_1 \propto (1-\bar \Sigma^1_1) ( 1+ \bar \Sigma^0_0+\bar \Sigma^2_2+\bar \Sigma^3_3)
%\ee
%Thus
Then the speed of propagation along the $x^1$ direction is given by
\be
c_1^2 = \frac{Z^1_{\ 1}}{Z^0_{\ 0}} = \frac{(1-\bar \Sigma^1_{\ 1}) ( 1+ \bar \Sigma^0_{\ 0}+\bar \Sigma^2_{\ 2}+\bar \Sigma^3_{\ 3})}{(1-\bar \Sigma^0_{\ 0}) ( 1+ \bar \Sigma^1_{\ 1}+\bar \Sigma^2_{\ 2}+\bar \Sigma^3_{\ 3})} \, .
\ee
Then whenever the `time' eigenvalue of $\bar \Pi$ and hence $\bar \Sigma$ is larger than the space eigenvalues, \ie  $\bar \Pi^0_{\ 0} > \bar \Pi^1_{\ 1}$ which implies $\bar \Sigma^0_{\ 0} > \bar \Sigma^1_{\ 1}$, the speed of propagation in that frame is larger than unity\footnote{For $\bar \Sigma^0_{\ 0}=1$ there is even classical instantaneous propagation but of course the regime for which one can trust this classical calculation is null in that case.}. This is both the low-energy group and phase velocity. However it is not the front velocity since quantum corrections are important when computing this quantity.

\subsubsection*{$\bullet$ Classical velocity in the $\rho$-duality frame}

On the other hand, in the $\rho$-duality frame, $\delta \rho$ is manifestly luminal, at least around backgrounds for which $\bar \chi(x)=\bar{\tilde{\chi}}(\tilde{x})=0$, from \eqref{tilde E pi} we simply have
\ba
\Box \delta \rho(x) =0 \,.
\ea
In addition $\delta \tilde \chi$ travels in an effective metric $\gamma_{\mu\nu}$ for which
\ba
\gamma^{\mu \nu} \propto \eta^{ \alpha  \beta} [(1- \Sigma)^{-1} ]_{\ \alpha}^{\mu}  [(1- \Sigma)^{-1} ]_{\ \beta}^{\nu} \propto \eta^{ \alpha  \beta} [1+ \Pi ]_{\  \alpha}^{\mu}  [1+ \Pi ]_{\ \beta}^{\nu}\,.
\ea
For this metric, whenever the `time' eigenvalue of $\Pi$ (or $\Sigma$) is larger than the space eigenvalues then the speed of propagation is less than unity:
\ba
c_1^2 = \frac{\gamma^1_{\ 1}}{\gamma^0_{\ 0}} =\frac{(1+\bar \Pi^1_{\ 1})^2}{(1+\bar \Pi^0_{\ 0})^2}= \frac{(1-\bar \Sigma^0_{\ 0})^2}{(1-\bar \Sigma^1_{\ 1})^2}\,.
\ea

\subsection{Classical Causal Structure}

\begin{figure}[ht!]
\centering
\includegraphics[width=150mm]{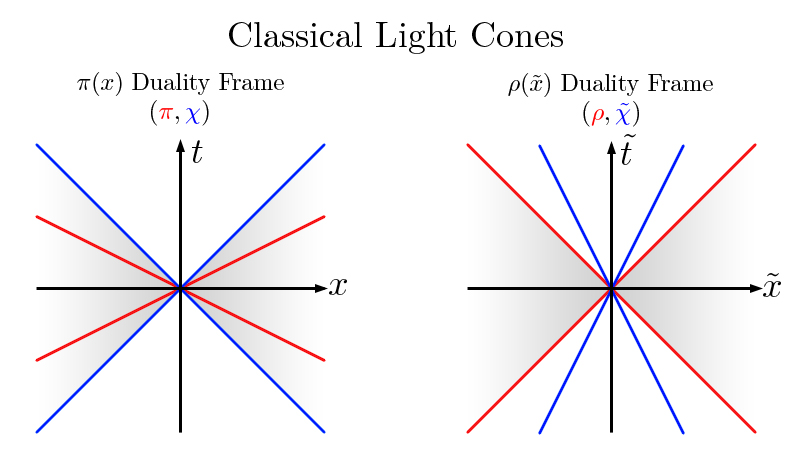}
\caption{Classical lightcones in the two duality frames for the example \eqref{EG} for background solutions for which $\bar \chi=0$ and $\bar \Sigma^0_{\ 0} > \bar \Sigma^1_{\ 1}$. The relative orientation is preserved even though the maximal speed is different. These lightcones are given in the local Lorentz frame for which $\bar \Pi$ and $\bar \Sigma$ are diagonal.}
\label{Pic:LightCone}%\vspace{10pt}
\end{figure}

The common feature in both frames is that for those solutions where $\bar \Sigma^0_{\ 0} > \bar \Sigma^1_{\ 1}$ then $\pi$ (resp. $\rho$) travels faster than $\chi$ (resp. $\tilde \chi$) and when $\bar \Sigma^0_{\ 0} < \bar \Sigma^1_{\ 1}$ then  $\pi$ (resp. $\rho$) travels slower than $\chi$ (resp. $\tilde \chi$). Consequently the lightcone structure is the same in both duality frames even though the limiting velocity is different (see Figure~\ref{Pic:LightCone}).\\

We can also find solutions in the $\rho$-duality frame for which the fluctuations of $\tilde \chi$ are superluminal (classically), \ie whenever $\bar \Sigma^0_{\ 0} < \bar \Sigma^1_{\ 1}$. However these map back into the $\pi$-duality frame as solutions for which $\chi$ is luminal and $\pi$ is subluminal. It is also true that we can find solutions for which there are classical superluminalities in both duality frames. For instance if $\bar \Sigma^0_{\ 0} > \bar \Sigma^1_{\ 1}$  and $\bar \Sigma^0_{\ 0} < \bar \Sigma^2_{\ 2}$ then matter travelling in the direction $x^1$ in the $\rho$-duality frame is subluminal, but it is superluminal in the direction $x^2$. An analogous inverse statement holds for $\pi$ in the $\pi$-duality frame.
The classical picture of the causal structure is given in Figure~\ref{Pic:LightCone}.

\begin{figure}[ht!]
\centering
\includegraphics[width=150mm]{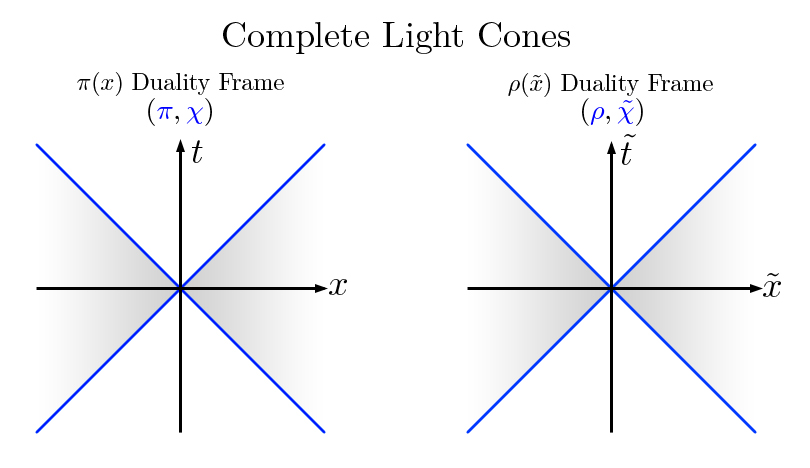}
\caption{Light cones in a UV description of Galileons including non-perturbative quantum corrections. In a UV completion the front velocity is always expected to be luminal regardless of the background.}
\label{Pic:LCQC}
\end{figure}

This result appears miraculous until we realize that it is just a reflection of the fact that the duality transformation acts effectively as a coordinate transformation. In the presence of a background field configuration $\bar \Pi$, then we can think of the duality transformation as a combination of a background coordinate transformation of the form
\be
\tilde x^{\mu} = x^{\mu} + \frac{s}{\Lambda^{\sigma}} \partial^{\mu} \bar \pi\,,
\ee
combined with a linearized field redefinition of $\delta \pi(x)$
\be
[\delta \rho](\tilde x) = [\delta \pi](x) \, .
\ee
If the matter fields travel at the speed of light in the original frame, then they travel on a metric determined by $\d s^2 = \eta_{\mu \nu} \d x^{\mu} \d x^{\nu}$. But by the background transformation this maps into the following metric
\be
\d s^2 = \eta_{\mu \nu} \d x^{\mu} \d x^{\nu} = \eta_{\mu \nu} [1-s \bar \Sigma]^{\mu}_{\ \alpha} [1-s \bar \Sigma]^{\nu}_{\ \beta} \d \tilde x^{\alpha} \d \tilde x^{\beta}\,,
\ee
which again manifests the property that in the new coordinate system whenever $\bar \Sigma^0_{\ 0} > \bar \Sigma^1_{\ 1}$ the matter travels subluminally.\\

To summarize, we do not view there to be a paradox here. One should not expect the velocities to be invariant under the duality. The fact that a velocity in one frame can appear superluminal and is luminal in another frame does not imply a discrepancy but rather a signal that these superluminalities cannot be used to infer causality or absence thereof. To reemphasize the two essential points:
\begin{itemize}
\item One does not expect the low energy group/phase velocity to be invariant under the duality. However we do expect the physics such as the causal structure to remain unchanged under the duality. The causal structure is not dictated by the group/phase velocity but rather by the front velocity.
\item The front velocity is the high frequency limit of the phase velocity. By its very definition quantum corrections need to be taken into account when computing it. Thus the previous results which were classical do not lead to the correct front velocity. The front velocity with its quantum corrections will be considered in a UV description of Galileons in \cite{Luke}.
\end{itemize}
%These two points are discussed further in what follows.

\subsection{Front velocity and importance of quantum corrections}

Let us now address the question of why we might expect the quantum front velocity to be different from the classical one. Certainly since perturbation theory breaks down at the scale $\Lambda$, we can expect order unity corrections to the front velocity at that scale. However by itself this is not an argument that luminality will be recovered. We will argue elsewhere that at least for a certain class of theories, there is a mechanism by which luminality can be recovered \cite{Luke}.
The key physics is the Vainshtein mechanism \cite{Vainshtein:1972sx} applied at the quantum level (see also \cite{deRham:2013qqa}). The theories that exhibit superluminal group velocities at low energies are precisely those theories that exhibit the Vainshtein mechanism\footnote{And the reverse is also true, \cite{Dvali:2012zc,Vikman:2012bx}: theories that exhibit classicalization also exhibit superluminal group velocity. }.  Associated with a quanta of frequency $\omega$ there is a Vainshtein radius (also called classicalization radius in \cite{Dvali:2010jz}), which is given by
\be
r_{\star} = \Lambda^{-1}\left( \frac{\omega}{\Lambda}\right)^{\frac{1}{(d+1)}} \, .
\ee
Consider now the evaluation of the two point function from the path integral
\be
\langle \alpha, {\rm out} | T{\cal O}(x) {\cal O}(y) | \alpha, {\rm in} \rangle = \int D \pi  e^{i S[\pi]} {\cal O}(x) {\cal O}(y) \, .
\ee
From this we may infer the front velocity by determining the behaviour of its Fourier transform at high $k$/high $\omega$. Now as $k$ and $\omega$ are increased, the path integral becomes dominated by its saddle point solutions because the Vainshtein mechanism forces the associated saddle point solution to grow in size since $r_{\star}$ is a positive power of $\omega$. The saddle-point action itself scales as $S \sim  \left( \frac{\omega}{\Lambda}\right)^{\frac{d+2}{(d+1)}}$ which grows as a positive power of $\omega$ confirming that the semi-classical approximation should become better and better\footnote{Note that this is not the same as the tree level approximation to the propagator becoming better. Here the semi-classical contributions are more akin to instanton $e^{-S_{\rm instanton}/\hbar}$ contributions which are still inherently quantum. Precisely these instanton contributions are considered in analogous system in \cite{Alberte:2012is}.}.\\

The result is that the Vainshtein mechanism creates a type of UV/IR mixing whereby the UV high energy behaviour is dominated by large size IR saddle point solutions (this is called classicalization in  \cite{Dvali:2010jz}, we shall find independent arguments in favor of this reasoning in \cite{Luke}). This UV/IR mixing means that the front velocity is determined not by the small distance behaviour of the classical solution, but rather by the large distance behavior. Thus for example if we consider background solutions which are asymptotically trivial $\pi \rightarrow 0$ as $r \rightarrow \infty$ then the front velocity is determined by the region of the theory which is asymptotically Lorentz invariant. This in turn implies that the front velocity is luminal, leading to the causal picture of Figure~\ref{Pic:LCQC}. The details of these arguments will be left to \cite{Luke} (see also \cite{Dvali:2012zc} for closely related arguments).

\section{Discussion}
\label{sec:discussion}

The existence of a Galileon duality which maps a Galileon theory to another Galileon theory was proposed in \cite{Curtright:2012gx,deRham:2013hsa}. Interestingly a free, manifestly UV complete, causal and Lorentz invariant theory was shown to map to a quintic Galileon which exhibits superluminal group velocity (classically) and irrelevant operators. These results are closely related to those of \cite{Creminelli:2013fxa} where a similar mapping occurs for the conformal Galileon. The two dualities can be understood as a twist in the representation of the coset for the Galileon algebra $Gal(3+1,1)/ISO(3,1)$ or conformal algebra $SO(4,2)/ISO(3,1)$ \cite{Creminelli:2014zxa}.

In this manuscript we have   generalized  the duality to an arbitrary Generalized Galileon which corresponds to the most general local, Lorentz invariant single field theory with no ghosts. We have further generalized the duality to arbitrary local couplings to matter and found that it is interpretable as a diffeomorphism in a gravitational theory. This leads to many important consequences:
\begin{itemize}
\item First the duality map fully preserves the notion of locality.
\item Second it gives a strong argument in favor of the validity of the duality at the quantum level since the duality could only be violated at the quantum level if  diffeomorphism anomalies were present.
\item  Recognizing the duality as a diffeomorphism allows for a simple vector generalization of the duality.
\item  Using the vector duality, we have presented the dual of the Maxwell theory. The theory is $U(1)$ invariant but in a non-manifest way as the $U(1)$ is non-linearly realized. In particular the theory is not built solely out of the Maxwell tensor. This might open the door for a new class of $U(1)$-invariant theories.
\item Using the vector duality,  we were able to derive the dual to a Proca theory and have presented specific types of interactions which only propagate three degrees of freedom in four dimensions.
\item Recognizing the duality as a diffeomorphism has allowed us to include arbitrary couplings to matter fields with arbitrary spins. We find that they have to transform as in a normal diffeomorphism. With this transformation, the duality maps a local coupling to a local coupling. The appearance of non-local coupling to matter as pointed out in \cite{deRham:2013hsa,Creminelli:2014zxa} is an artifact of using an external source.
\item We have included the coupling to gravity in a way which respects the duality and shown how massive gravity is a natural example of a gravitational theory which is invariant under the duality. The same is true for bi-- and multi--gravity.
\item We have proposed a new class of  massive gravity theories where the \stu fields may enter in different parameters of the theory. We argue that the theory is local, Lorentz invariant, free of the Boulware--Deser ghost and has only five dynamical degrees of freedom about any background in the gravitational sector.  The linearized theory around the vacuum breaks translation invariance and therefore does not take the usual Fierz--Pauli form, but these theories are nonetheless Lorentz invariant. The same generalization can be made to bi-- or multi--gravity.
\item Finally this duality allows us to shed light on the classical superluminality and causality discussion in Galileon theories. We show that the superluminality can map to (sub)luminal propagation. Nevertheless the causal structure remains unchanged at the classical level. We emphasize that these classical considerations are not sufficient to correctly compute the front velocity which is the quantity that needs to be luminal to ensure causality.  In further work \cite{Luke} the front velocity is computed non-perturbatively and shown to be unity which if correct would imply that Galileon theories can be causal despite exhibiting superluminal classical group velocity. This  would be in full agreement with results derived from the duality.
\end{itemize}

To summarize this manuscript has derived a multitude of results relying on the classical duality and strong indications that the duality can be used at the quantum level. Results at the quantum level are beyond the scope of this manuscript and will be presented in   \cite{Luke}.

\acknowledgments

 CdR is supported by a Department of Energy grant DE-SC0009946. AJT is supported by Department of Energy Early Career Award DE-SC0010600.  AJT would like to thank the Perimeter Institute for Theoretical Physics for hospitality while this work was being completed. We would like to thank Paolo Creminelli, Matteo Fasiello, Austin Joyce, Andrew Matas, Nick Ondo, Marco Serone, Gabriele Trevisan and Enrico Trincherini for useful discussions and comments on the manuscript.

\appendix

\section{Duality for Non-Lorentz Invariant Systems}

\label{sec:appendix}

The duality map we have described in the main text manifestly preserves Lorentz invariance. If we are willing to give up manifest Lorentz invariance, the duality map can be extended to a $d(d+1)/2$ parameter tensor abelian group of transformations given by
\ba
&& \tilde x^{\mu} = x^{\mu} + \frac{s^{\mu\nu}}{\Lambda^{\sigma}}\partial_{\nu} \pi(x) \,  ,\\
&& x^{\mu} = \tilde x^{\mu} - \frac{s^{\mu\nu}}{\Lambda^{\sigma}}\tilde \partial_{\nu} \rho(\tilde x) \, .
\ea
Where now $s^{\mu\nu}$ is a constant symmetric tensor. These relations can be combined with
\ba
&& \pi(x) = \rho(\tilde x) -\frac{s^{\mu\nu}}{2 \Lambda^{\sigma}} \tilde \partial_{\mu} \rho(\tilde x)\tilde \partial_{\nu} \rho(\tilde x)  \,, \\
&& \rho(\tilde x) = \pi( x) + \frac{s^{\mu\nu}}{2 \Lambda^{\sigma}} \partial_{\mu} \pi(x) \partial_{\nu}  \pi(x) \, .
\ea
As before we have $\partial_{\mu} \pi(x) = \tilde \partial_{\mu} \rho(\tilde x)$. Now provided we define $\Pi_{\mu\nu}(x) = \partial_{\mu} \partial_{\nu} \pi(x)/\Lambda^\sigma$ and $\Sigma_{\mu\nu}(\tilde x) = \tilde \partial_{\mu} \tilde \partial_{\nu} \rho(\tilde x)/\Lambda^\sigma$ as before then we have
\be
\left( \Sigma^{-1} \right)^{\mu\nu}(\tilde x) = \left( \Pi^{-1} \right)^{\mu\nu}(x) + s^{\mu\nu} \, .
\ee
In particular we may choose $s^{0 \mu}=0$ and $s^{ij}= s \delta^{ij}$ for $i,j = 1,\ldots, d-1$ so that the duality acts only in the space directions. This may useful in studying non-relativistic theories.

 %\vspace{-30pt}

\bibliography{refs}

\end{document}